\documentclass[%
 preprint,
 amsmath,amssymb,
]{revtex4-1}

\usepackage{graphicx}
\usepackage{amssymb}
\usepackage{setspace}
\usepackage[utf8]{inputenc}
\usepackage{array}
\usepackage{natbib}
\usepackage{amsmath}
\usepackage{multirow}
\usepackage{hyperref}
\usepackage{subfigure}

\usepackage{color}
\definecolor{redcolor}{rgb}{1.0,0.,0.}

\definecolor{bluecolor}{rgb}{0,0.,1}

\begin{document}

\preprint{}

\title{Associations between author-level metrics in subsequent time periods}

\author{Ana C. M. Brito$^1$,  Filipi N. Silva$^2$ and Diego R. Amancio$^1$}

\affiliation{$^1$Institute of Mathematics and Computer Science, University of S\~ao Paulo, S\~ao Carlos, Brazil\\
$^2$Indiana University Network Science Institute, Bloomington, Indiana 47408, USA\\
}

\date{\today}

\begin{abstract}
%
Understanding the dynamics of authors is relevant to predict and quantify performance in science.
While the relationship between recent and future citation counts is well-known, many relationships between scholarly metrics at the author-level remain unknown. In this context, we performed an analysis of author-level metrics extracted from subsequent periods, focusing on visibility, productivity and interdisciplinarity. First, we investigated how metrics controlled by the authors (such as references diversity and productivity) affect their visibility and citation diversity. We also explore the relation between authors' interdisciplinarity and citation counts. The analysis in a subset of Physics papers revealed that there is no strong correlation between authors' productivity and future visibility for most of the authors. A higher fraction of strong positive correlations though was found for those with a lower number of publications. We also found that reference diversity computed at the author-level may impact positively authors' future visibility. The analysis of metrics impacting future interdisciplinarity suggests that productivity may play a role only for low productivity authors. We also found a surprisingly strong positive correlation between references diversity and interdisciplinarity, suggesting that an increase in diverse citing behavior may be related to a future increase in authors interdisciplinarity. Finally, interdisciplinarity and visibility were found to be moderated positively associated:  significant positive correlations were observed for 30\% of authors with lower productivity.
%
%

\end{abstract}

\maketitle



\section{Introduction}


The age of information promoted several new discoveries in science, with many of them emerging from interdisciplinary endeavors~\cite{lariviere2015long,carusi2019look,silva2013quantifying}. At the same time, these communities are growing in size and productivity, resulting in an ever-increasing deluge of digital information available in the form of published articles~\cite{pan2018memory}, datasets, and algorithms, as well as across many platforms, such as cloud services and social media. However, the increase in digital resources has not leveled the playing field for researchers; inequality is rising in science~\cite{xie2014undemocracy}.

Understanding the mechanisms leading to inequality in science can help policymakers and funding agencies better distribute research resources while also promoting a more just and democratic environment. Part of this problem relies on the fact that researchers compete among themselves for limited funding and attention. In such a system, an increase of researchers’ visibility leads to better funding opportunities which, in turn, leads to more availability of resources for their institutions, thus allowing those researchers to attain even greater visibility.

The cycle in which researchers with the most resources are rewarded with even more resources over time is a source of inequality known as the Matthew Effect~\cite{merton1968matthew}.
This is one of the reasons why understanding how the dynamics of authors visibility unfold over time is one of the most important problems in the field of Science of Science~\cite{fortunato2018science,sinatra2016quantifying}. However, not much attention was given to understand the relationships between other authors metrics besides citations~\cite{ren2012modeling,eom2011characterizing,amancio2012three,simkin2005stochastic}. In particular, the literature lacks studies on metrics controlled by the authors, such as those based on their choices of references and their productivity; and the possible effects they may have on their received citations.

Here, we propose to explore the associations among different bibliometric measures for authors in subsequent time periods through correlation. Among the metrics we consider are interdisciplinarity, which is measured in terms of the subject diversity~\cite{silva2013quantifying} of citations received by the authors, productivity and visibility of authors. Here, we are interested in addressing three main questions:
\begin{enumerate}
    \item How metrics controlled by the authors -- namely, their productivity and diversity in the choice of references -- correlate with visibility metrics, such as the received number of citations per paper, in a subsequent time period?
    \item How these characteristics correlate with the future interdisciplinarity of their publications based on citations?
    \item How interdisciplinarity is related to future citations and vice versa?
\end{enumerate}

In addition to productivity and visibility, we also studied interdisciplinarity as it plays an important role in modern science given the increasing number of authors bridging new different fields. Here, it is used as a descriptor for citation diversity, as adopted in related works~\cite{silva2013quantifying}. In a similar fashion, we adopt a reference diversity for authors based on the fields of the employed references in their publications.
%


We employ the  \emph{American Physical Society} (APS) dataset, which incorporates all the citations and metadata for papers, mainly in Physics, published in any of the APS journals up to 2010. More specifically, we employ the dataset used in~\cite{recency}, which was supplemented with disambiguated authors from the \emph{Microsoft Academic Graph} (MAG). First, we construct a co-occurrence network for the categories existing in the APS journals (PACS codes) which is used to define a metric of interdisciplinarity for authors based, in terms of the diversity of their received citations or the references they used. Next, we calculate the correlation between the considered author-level metrics for a window considering previous publications and another which considers subsequent publications and citations. Finally, we use a statistical framework based on null models to obtain the significance of the correlations between the considered metrics.






Several interesting results have been obtained in our analysis.
We found that the diversity of references may impact positively the observed future visibility for 1/3 of low-productivity authors. This effect is minimized when analyzing more productive authors, yet the fraction of authors that were positively affected varied between $22\%$ and $25\%$. A weaker association between productivity and citation counts was found: the highest fraction of authors with a significant positive correlation was $21\%$.
When comparing the fraction of authors displaying significant positive and negative correlations, both productivity and reference diversity turned out to be more positively than negatively correlated with authors' visibility.
Surprisingly, we also found that reference diversity and future interdisciplinarity are strongly positively correlated for roughly $50\%$ of authors. Finally, the association between interdisciplinarity and visibility revealed that an increase in interdisciplinarity is more likely to be linked to an increase in visibility for low productivity authors. Such positive significant correlations were observed in roughly $30\%$ of authors in that class.
We believe our results can provide further insights into better understanding researchers' career dynamics.

\section{Related works}


In this paper, among other relationships, we analyze which factors affect the visibility of authors (measured in terms of citations). At the paper level, some correlations between paper features and the number of citations have been studied in the last few years. An important factor that has been found to affect the visibility of paper is related to the \emph{interdisciplinarity} of venues in which they are disseminated. Different aspects of scientific pieces have been used to define interdisciplinarity indexes. In~\cite{silva2013quantifying}, journal citation networks are used to quantify how interdisciplinary a journal is. For a given journal, the diversity of citations from different areas is used to gauge interdisciplinary. Such a diversity is computed using the concept of \emph{true diversity}, a measure widely used to express how diverse a set of elements from different classes is~\cite{tuomisto2010diversity,correa2017patterns,tohalino2018extractive}. Subject areas and citation data were extracted from the \textit{Journal Citation Reports} dataset. Some interesting conclusions were the positive correlation between the proposed interdisciplinary index and journals impact factor. In other words, interdisciplinary journals tend to have a higher impact factor than specialized journals.

Using a different approach, the study conducted in~\cite{carusi2019look} also quantified journals interdisciplinarity. The authors used Scopus data comprising \emph{Information and Communication Technology} publications. The relationship between scholars and journals was represented via bipartite graphs. After a SVD dimension reduction, a spectral co-clustering method was used to identify communities of scholars and journals.
The diversity (i.e. the interdisciplinarity) of a journal was then defined by analyzing the unevenness of authors distribution over the obtained network communities. Such a dispersion was computed via Shannon entropy, Simpson diversity, and Rao-Stirling index~\cite{leydesdorff2019interdisciplinarity}. High values of disparity metrics were found to occur in journals appearing between communities. Conversely, low diversity was observed mostly in network community cores. 

%

%



A correlation between interdisciplinarity and citation impact was investigated in~\cite{yegros2015does}. Three aspects of interdisciplinary were investigated at the paper level: variety, balance, and disparity. Variety is the total number of different disciplines (or \emph{Web of Science} categories) cited by the paper, while balance corresponds to the evenness of the disciplines distribution, computed via Shannon diversity. Disparity measures how different are the disciplines in the reference set. The authors analyzed the impact of papers using the Normalized Citation Score (NCS).
The data set used was papers from Science Citation Index-Expanded (2005).
A regression estimation analysis revealed that variety was positively associated with NCS. In contrast, both balance and disparity were negatively associated with NCS.

The impact of citing interdisciplinary papers on papers visibility was investigated in~\cite{lariviere2015long}. The authors characterized interdisciplinarity at the paper level by using papers references. Subdisciplines were defined by the UCSD map of science~\cite{borner2012design}. {According to this map, the similarity between journals is based on the number of shared references (via bibliographic coupling) and keywords. An average-linkage clustering strategy generates a cluster of $13$ different categories and the pairwise cluster distance is represented in a 3D Fruchterman-Reingold layout.}
%
An analysis of 25,000 documents showed that papers citing interdisciplinary sub-disciplines tend to receive more citations than papers with fewer references to interdisciplinary sub-disciplines. This study also grouped sub-disciplines by distance in the UCSD map and demonstrated that papers citing distant sub-disciplines tend to have higher relative citation rates than papers citing similar sub-disciplines.
%

%
%
%

At the author level, the study carried out in~\cite{pluchino2019exploring} investigated the effects of interdisciplinarity on scientists careers.
The APS dataset was used, considering papers published between 1980 and 2009. The hierarchical system of subdisciplines classification -- referred to as \emph{Physics and Astronomy Classification Scheme} (PACS) -- was used to measure the interdisciplinarity of an author. They proposed an index combining the total of PACS codes used during the entire author career and the average number of different classes appearing simultaneously in the author papers. Using this value, authors were grouped by different levels of interdisciplinarity: low, medium, and high. Based on these groups, it was observed that higher interdisciplinarity affects positively productivity. A statistical model was proposed to reproduce the original data. The factors considered in the model were the proposed interdisciplinarity index, the number of publications in each class, the number of citations, talent, reputation, and luck. 
The model reproducing the properties of the studied system revealed that authors with medium-high talent are the most successful ones. In addition, luck turned out to play an important role in career success. Surprisingly, it was found to be even more relevant than interdisciplinarity factors in some cases.

Another different source of factor concerns the well-known rich-get-richer paradigm. In other words, if an author has received several citations, he/she has a higher tendency of receiving more citations if they have received a higher citation rate in the past. In~\cite{recency}, the authors describe a model for reproducing the distribution of authors citations in the APS dataset. Unlike other models, they included a recency factor so that more recent citation data receives a higher weight in the preferential attachment model. This model showed that the rich-get-richer paradigm describes the citation distribution for authors publishing in APS journals. Most importantly, they also found that recency plays an important role to define how broad the burstiness of citations are. The number of citations received by authors is strongly dependent on the total of citations received in the last 1-2 years~\cite{recency}.

\section{Methodology} \label{sec:methodology}

The methodology adopted in this paper can be divided into the following steps:

\begin{enumerate}

    \item \emph{Creation of PACS networks}: this phase is responsible for establishing and identifying the subfields inside the considered dataset. Groups of strongly connected subareas are grouped into network communities. The latter is used to identify an area, which in turn is used to define some of the variables of interest. The dataset used to create the networks is described in Section \ref{sec:data}. The process of creating and identifying communities of co-occurring PACS is described in Section \ref{sec:panet}.

    \item \emph{Definition of diversity indexes}: here we use diversity indexes to quantify how diverse authors cite or are cited by other papers. The diversity takes as reference the subareas (communities) identified in the PACS network. The adopted diversity index is defined in Section \ref{sec:div}. Diversity indexes are among the author-level metrics of interest in this paper.

    \item \emph{Quantifying the relationship between variables of interest}: here we quantify there are correlations between variables of interest quantified in subsequent time intervals. The methodology adopted to quantify the fraction of authors displaying significant positive/negative correlations between the variables of interest in described in Section \ref{sec:pastfut}.

\end{enumerate}

\subsection{Dataset} \label{sec:data}


The dataset consists of papers published by the American Physical Society (APS) journals between 1991 and 2010. The dataset comprises 299,930 publications from APS journals.
While the dataset provides several article metadata, we used for each paper the list of authors and the reference list. We also used the list of subfields codes provided by the authors and selected from the \emph{Physics and Astronomy Classification Scheme} (PACS). This classification scheme is a hierarchical code system used to organize the main fields and subfields in Physics journals.

When addressing any issue at the author level, one should be aware that ambiguities and name split may arise~\cite{amancio2012use,milojevic2013accuracy}. To address this problem, we used the Microsoft Academic Graph (MAG) dataset, which is a more extensive set of publications with authors' names disambiguated~\cite{recency}. We mapped the APS dataset into the MAG database by matching DOIs values.
%

\subsection{PACS Networks} \label{sec:panet}
\label{subsec;pacs_nets}





\begin{figure}[ht!]
    \centering
    \includegraphics[width=0.70\textwidth]{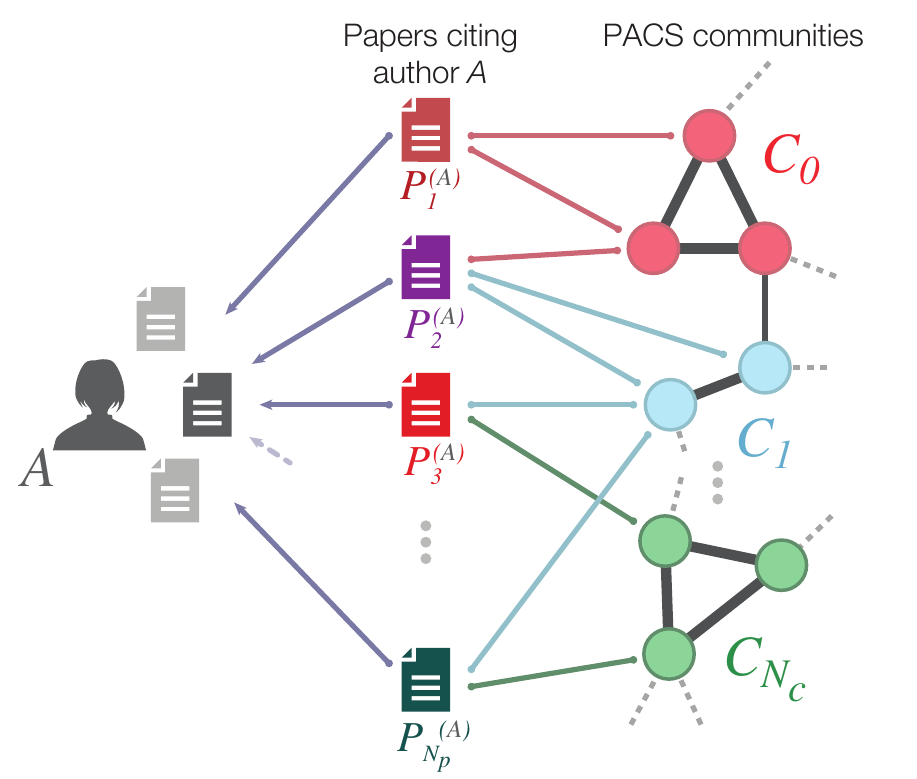}
    \caption{Schematic representation of the components needed to calculate \emph{citations} and \emph{references} diversity.}
    \label{fig:schematic}
\end{figure}

In this work, we use the notion of subfields to compute the degree of interdisciplinarity inside the Physics area (for APS journals).
Subfields were derived from PACS co-occurrence networks~\cite{pan2012evolution}. Each publication in the APS dataset has its PACS codes, and this information of area is provided by the authors, among a list of possible codes. We used this information to generate networks where nodes are PACS codes. Figure \ref{fig:schematic} shows an example of PACS co-occurrence network extracted from a set of papers.  As suggested by other works, PACS were analyzed at the first two levels~\cite{pan2012evolution}. Two codes are linked whenever they appear together in one or more papers. Here we take the view that a subfield in the considered subset of Physics papers can be seen as a subset of highly connected codes. In this way, each subfield is defined as a community in the respective co-occurrence PACS network. While our results are based on the Louvain community detection algorithm~\cite{traag2019louvain}, a preliminary analysis revealed that there is no large difference when other methods are used to detect communities. Considering the most recent years of the dataset, using the Louvain method, we found $10$ network communities.
An analysis of the obtained communities considering data from the last 5 years showed that the four largest communities are mainly composed of papers in the following subjects: %
(i) \emph{magnetic properties and materials}; (ii) \emph{quantum mechanics, fiel theories, and special relativity}; (iii) \emph{structure of solids and liquids; crystallography}; and (iv) \emph{statistical physics, thermodynamics, and nonlinear dynamical systems}.



%

\subsection{Diversity indexes} \label{sec:div}


Here we employ a diversity index for authors based on the diversity of fields being cited (\emph{citations diversity}) or referenced (\emph{references diversity}) by their papers.
Because usually \emph{citation diversity} is related to \emph{interdisciplinary}~\cite{silva2013quantifying}, we use both terms to describe the same concept.
To assign a distribution of fields of a given author $A$, first, we look at all the papers $P^{(A)}_i$ citing publications co-authored by $A$ during the considered time window. For each citing paper we obtain the communities associated to the PACS listed in the paper. Figure~\ref{fig:schematic} illustrates the necessary components employed to calculate the \emph{in}-diversity index for authors. Next, we derive the weights $w_\mathrm{in}(P_i,C_j)$ relating a paper $P_i$ to a PACS community $C_j$, defined as the ratio of the number of PACS in $C_j$ listed in $P_i$, i.e.
\begin{equation}
    w(P_i,C_j) = {|\mathrm{PACS}(P_i) \cap C_j| \over |\mathrm{PACS}(P_i)|},
    \label{eq:wpaper}
\end{equation}
where $\mathrm{PACS}(P_i)$ is the set of PACS listed in paper $P_i$. Next, we assign a weight $\bar w_\mathrm{cit}(A,C_j)$ relating an author $A$ to each PACS communities $C_j$ based on the citing papers. Each citation to a paper from author $A$ counts as a unit that is distributed among the communities, so that $\bar w(A,C_j)$ is defined as
\begin{equation}
    \bar w_\mathrm{cit}(A,C_j) = \sum_{P_i} {n_\mathrm{cit}(P_i,A) w(P_i,C_j)},
    \label{eq:wauthorin}
\end{equation}
where $n_\mathrm{cit}(P_i,A)$ is the number of citations from $P_i$ to author $A$ 
Finally, we normalize $\bar w(A,C_j)$ across all the received citations, thus obtaining a probability-like measure $p_\mathrm{cit}(A,C_j)$ of relatedness between an author $A$ and a community $C_j$, given by
\begin{equation}     \label{eq:pauthorin}
    p_\mathrm{cit}(A,C_j) = {\bar w_\mathrm{in}(A,C_j) \over \sum_{C_k}\bar w_\mathrm{in}(A,C_k)}.
\end{equation}
The \emph{citation} diversity index $\mathrm{citDiv}(A)$ is then defined as the exponential of entropy of $p_\mathrm{cit}(A,C_j)$~\cite{silva2013quantifying}, i.e.,
\begin{equation}
    \mathrm{citDiv}(A) = \exp\Big{[}-\sum\limits_{C_j} p_\mathrm{cit}(A,C_j)\log p_\mathrm{cit}(A,C_j) \Big{]}.
    \label{eq:indiv}
\end{equation}
Similarly, to obtain \emph{references} diversity index, we use the papers $P_i$ referenced by works authored by author $A$ instead of the received citations. Thus, the weight linking an author and a PACS community is defined as
\begin{equation}
    \bar w_\mathrm{ref}(A,C_j) = \sum_{P_i} {n_\mathrm{ref}(A,P_i) w(P_i,C_j)},
    \label{eq:wauthorout}
\end{equation}
where $n_\mathrm{ref}(A,P_i)$ is the number of times author $A$ cited the paper $P_i$.
The probability analogous to $p_\mathrm{cit}$ (i.e. $p_\mathrm{ref}$) is then normalized as:
\begin{equation} 
    p_\mathrm{ref}(A,C_j) = {\bar w_\mathrm{ref}(A,C_j) \over \sum_{C_k}\bar w_\mathrm{ref}(A,C_k)},
    \label{eq:pauthorout}
\end{equation}
and the \emph{references} diversity $\mathrm{refDiv}(A)$ is calculated as
\begin{equation}
    \mathrm{refDiv}(A) = \exp \Big{[} -\sum\limits_{C_j} p_\mathrm{ref}(A,C_j)\log p_\mathrm{ref}(A,C_j) \Big{]}.
    \label{eq:indiv}
\end{equation}

Both equations \ref{eq:pauthorin} and \ref{eq:pauthorout} have been used to measure diversity in many contexts~\cite{silva2013quantifying,correa2017patterns,de2016using}. Because the computation of $p_\mathrm{cit}$ and $ p_\mathrm{cit}$ are not reliable when only a few data is available, these quantities were computed for authors with more than ten references and citations in the dataset.
%

\subsection{Past and Future scholarly time series} \label{sec:pastfut}

\begin{figure}[h]
    \centering
    \includegraphics[width=0.99\textwidth]{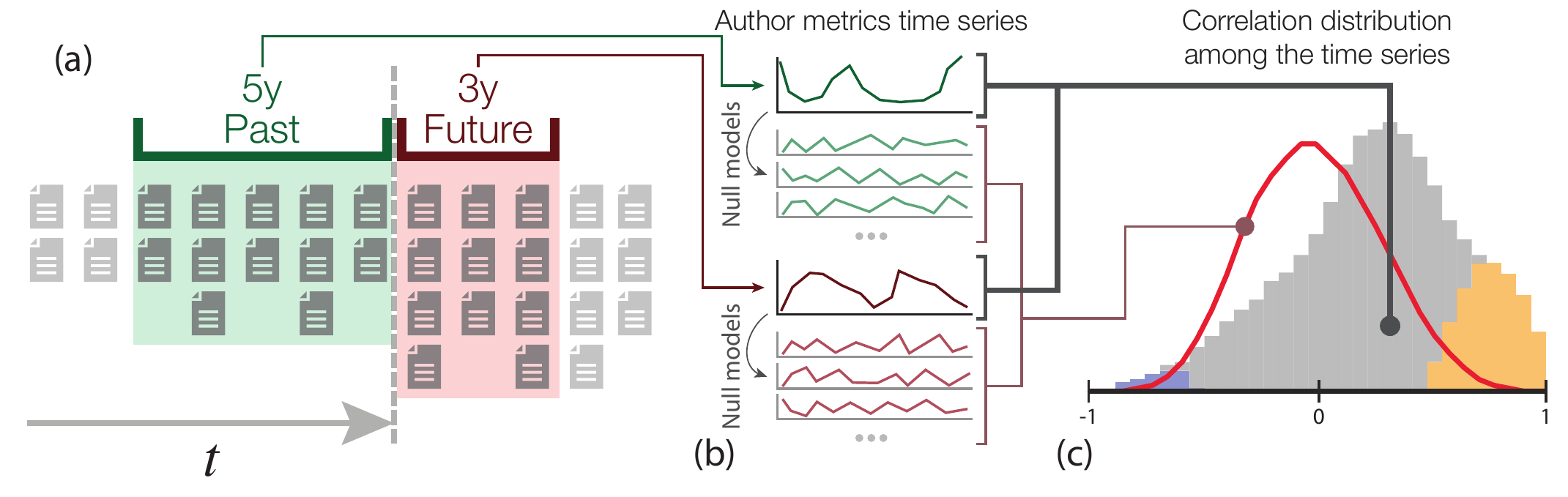}
    \caption{Schematic representation of the proposed methodology. (a) Given two subsequent windows (past and future) that moves over time, we calculate the time series of the considered metrics. (b) For each time series we derive a null model based on shuffling them along time. (c) We draw the correlation distribution (gray) obtained from the data time series and highlight negative (blue) and positive (yellow) values that are significant in comparison to the null models. The average null model distribution is also shown for comparison in red.}
    \label{fig:window}
\end{figure}

We propose a framework to analyze how a scholarly metric or diversity at a certain point in time for an author $A$ may impact his future metrics. First, we define two moving windows, one for the past and another for the future, respectively a 5 years window before the time under consideration $t$, and a 3 years window after $t$, as illustrated in Figure~\ref{fig:window}a. For each window, we calculate the scholarly metrics of $A$. In particular, for the Past window, we calculate the number of papers, citations received per paper, and references diversity, only considering publications in the period.

For the Future window, we calculate the number of citations received in that window from papers published by $A$ during the Past window. In the same fashion, we calculate citation diversity by considering only publications in the Past windows and citations in the Future window. By moving the windows along $t$ for a period from 1995 to 2010,
we obtain Past (number of papers, citations per author, and the references diversity), and Future (citations received per paper and citation diversity) time series for each author based on the calculated scholarly metrics.

In order to draw relationships between the scholarly metrics from past and future windows, we adopted the Pearson correlation. However, as these metrics may have characteristics that can lead to spurious correlations, such as the presence of outliers or long-tail distributions, we employed a statistical approach based to measure the significance of the obtained correlations. First, for each time series of each author, we obtain a set of $10,000$ surrogates generated by shuffling the original data along time, which is regarded as a null model, as illustrated in Figure~\ref{fig:window}b.

When calculating the correlations between two scholarly metrics, we also calculate the respective correlation distribution from their null models. This distribution is used to calculate a $p$-value associated with each author and a pair of past and future metrics. The $p$-value is defined as the probability of the null model resulting in a \emph{absolute} correlation that is higher than what was found for the data. Finally, the results are presented in the form of a correlation distribution alongside the percentage of negative and positive significant relationships by considering a threshold of $5 \times 10^{-2}$ for the $p$-values. This is illustrated in Figure~\ref{fig:window}c.

\section{Results and discussion}

Here we analyze the relationship between relevant author-level metrics.
More specifically, we analyze, if the diversity of references, the numbers of papers, and the number of references are correlated with citation counts and citation diversity.
We first focus on the relationship between variables that authors can control in the first 5-year window (e.g. the number and diversity of references) and variables that are not directly self-dependent (such as the number of citations and citation diversity) and are measured in the following 3-year window.
The correlations between paper/reference features and citation counts are discussed in Section \ref{sec1}.
The correlations between paper/reference features and citation diversity are discussed in Section \ref{sec2}.
Because interesting relationships between interdisciplinarity (i.e. citation diversity) and citation counts have been reported at different levels~\cite{silva2013quantifying,okamura2019interdisciplinarity,carusi2019look}, we also analyzed the correlations between interdisciplinarity and citations at the author level. This is reported in Section \ref{sec:interplay}.
%

\subsection{Correlations between reference features and citations} \label{sec1}


The simplest reference feature that can be used in our analysis is the total number of references. For the sake of clarity, we will use instead of the number of papers (i.e. the authors productivity) in our analysis because the total number of references is strongly related to the number of papers. In addition, the results using either number of references or the number of papers are very similar.

\begin{figure}[ht!]
    \subfigure[]{\includegraphics[width=0.49\textwidth]{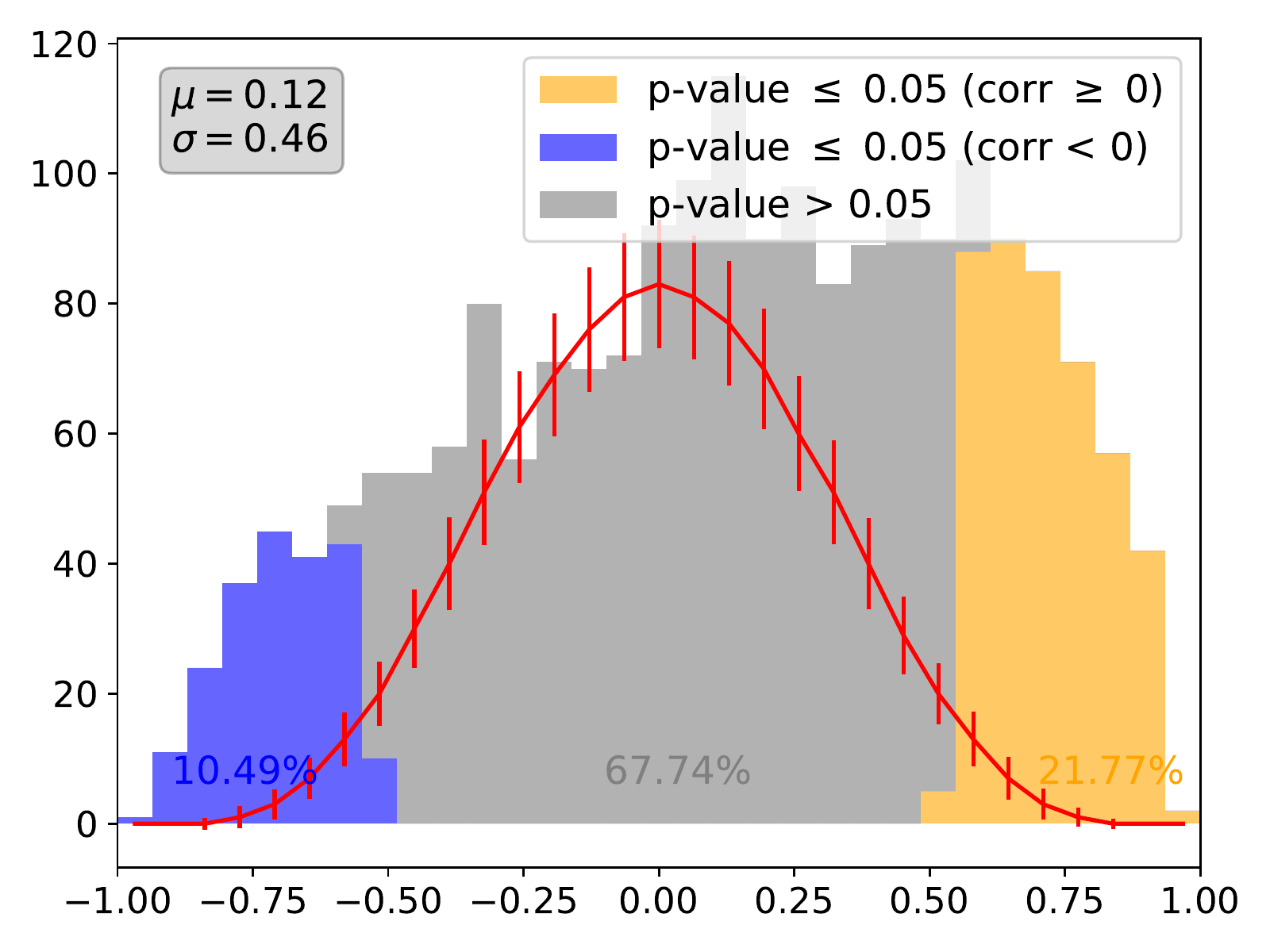}}
    \subfigure[]{\includegraphics[width=0.49\textwidth]{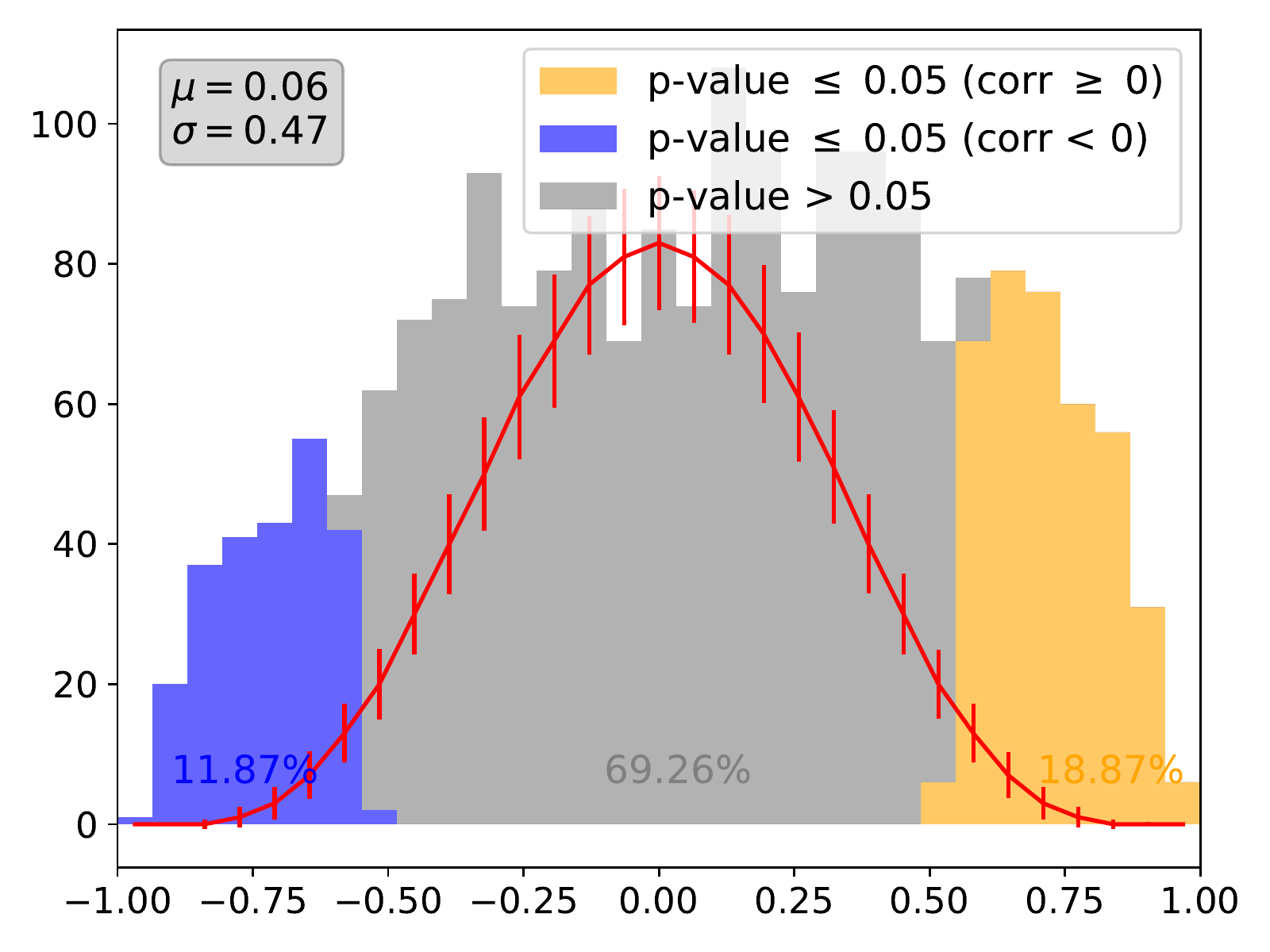}}
    \subfigure[]{\includegraphics[width=0.49\textwidth]{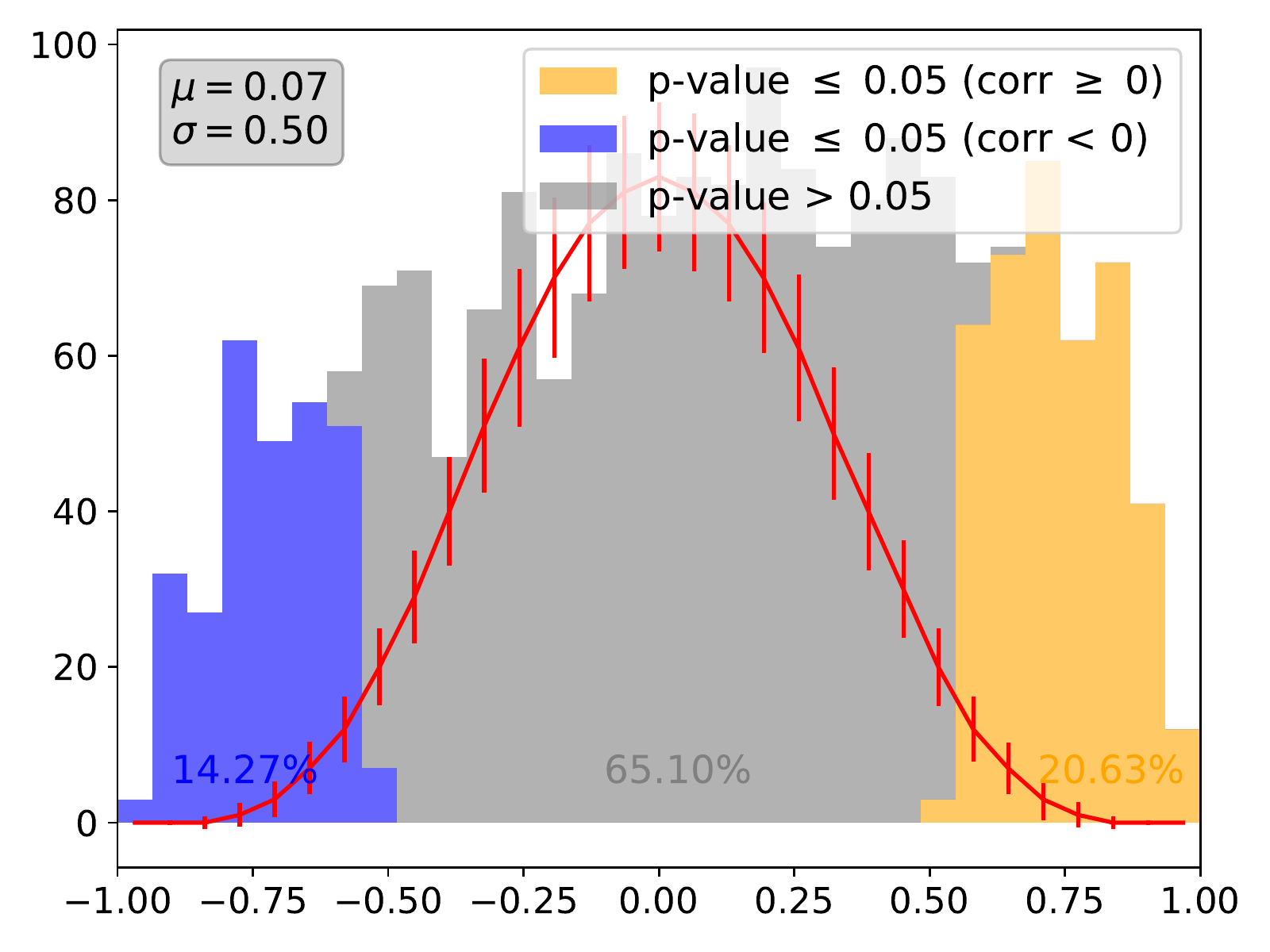}}
    \subfigure[]{\includegraphics[width=0.49\textwidth]{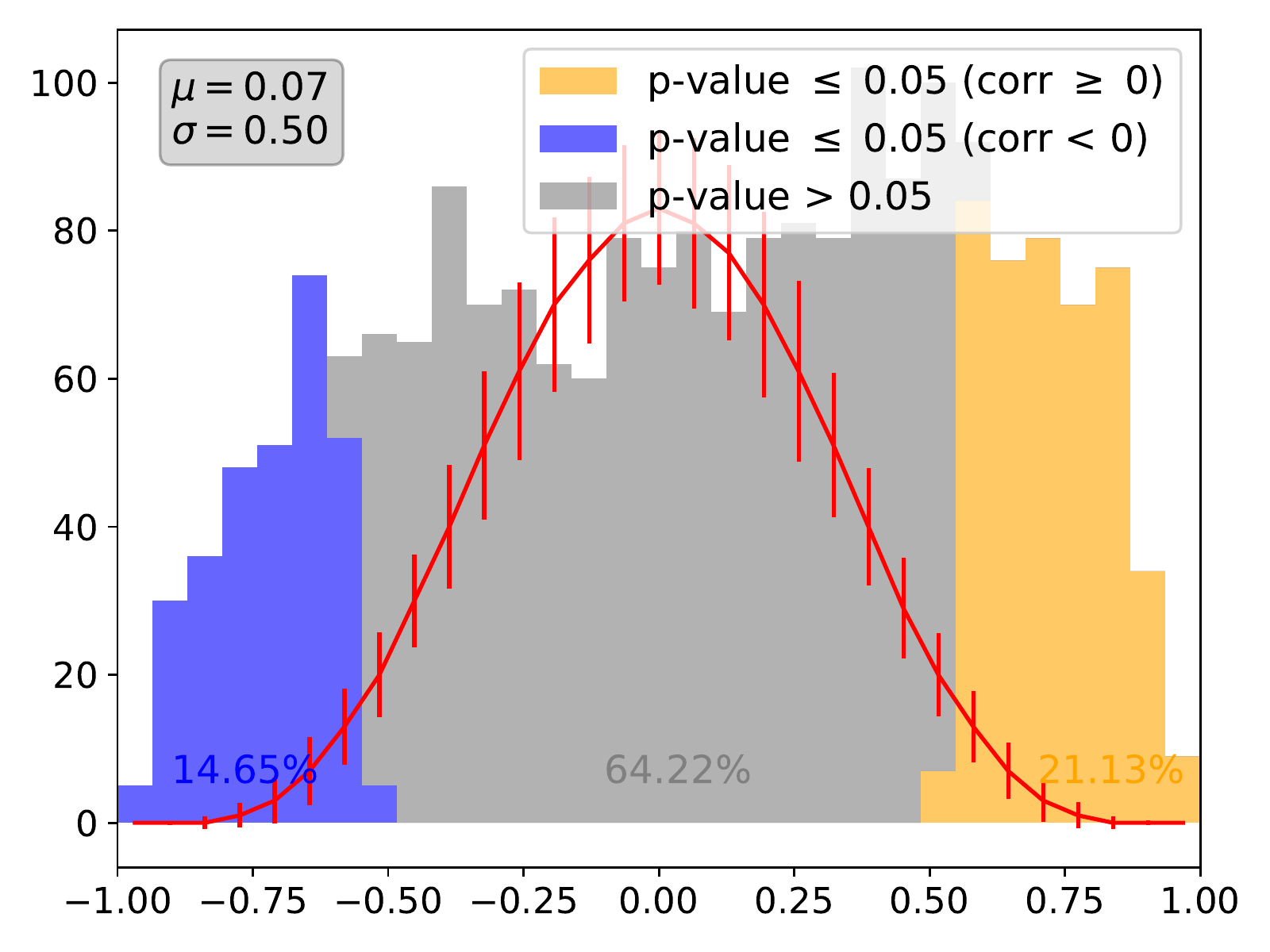}}
    \caption{
    Correlation between the \emph{total number of published papers} and \emph{citations per paper}. Panels (a)-(d) correspond to quartiles of authors sorted, in increasing order, by number of publications. The distribution of correlations obtained with the adopted null model is shown in the red curve.}
    \label{fig:pap-cit}
\end{figure}

We start our analysis by analyzing whether productivity -- i.e. the number of published papers -- is correlated with the total number of citations per paper.
This result is shown in Figure \ref{fig:pap-cit}. {As mentioned in the methodology (see Section \ref{sec:pastfut}),} the histograms shows the distribution of authors in different degrees of correlation between the variables of interest. Each panel corresponds to a different class of authors, according to its productivity. The authors analyzed in subpanels (a)-(d) are those who published the following amount of papers over all the considered period: (a) 5--25; (b) 26--36; (c) 37--58; and (d) 59--359 papers. The considered thresholds in the number of publications were chosen so that each class comprises 25\% of all authors in the dataset. In other words, each panel corresponds to a quartile of authors. In this figure, the distribution of correlations observed using the null model is represented by the red curve (see Section \ref{sec:pastfut}). The fraction of authors displaying significant positive and negative correlations between the considered variables are represented in yellow and blue, respectively.

The results in Figure \ref{fig:pap-cit} reveals that the observed distribution in all panels differs from the null model distribution. The discrepancy between real data and null model arises since very high or low values of correlations are unlikely to happen by chance, while the real data reveals an opposite effect: for a fraction of authors, the correlations are significant. Considering all four classes, 18-22\% of authors displayed a \emph{positive} correlation between productivity and visibility. On the other hand, a \emph{negative} correlation was also observed in all classes of authors. The percentage of authors displaying a \emph{negative} correlation between productivity and visibility ranged between 10\% and 15\%. Because more than 64\% of the observed correlations are not significant in all four classes of authors, the results suggest that for most of the authors the increase in productivity is not correlated with higher citation counts per paper.

In our analysis, we also compared the proportion of positive ($f^+$) and negative ($f^-$) correlations. The proportions are compared via $q$-index, defined as
\begin{equation}
    q = \frac{f^+}{f^-}. 
\end{equation}
In this case, all values of $q$ are higher than $q=1$, suggesting thus that in all classes of authors positive correlations are more likely to appear. The highest value of $q$ was observed for authors with the lowest number of publications (see panel (a)). We found $q=2.08$, meaning that positive correlations are twice more likely to appear than negative correlations considering this class of authors.

The results regarding productivity suggest that significant correlations between productivity and citation rates occur only to a small percentage of authors. This effect is more prominent in authors with lower productivity. This is an indication that, for most of the authors, increasing productivity does not improve authors' visibility in the near future.

\begin{figure}[ht!]
    \centering
    \subfigure[]{\includegraphics[width=0.49\textwidth]{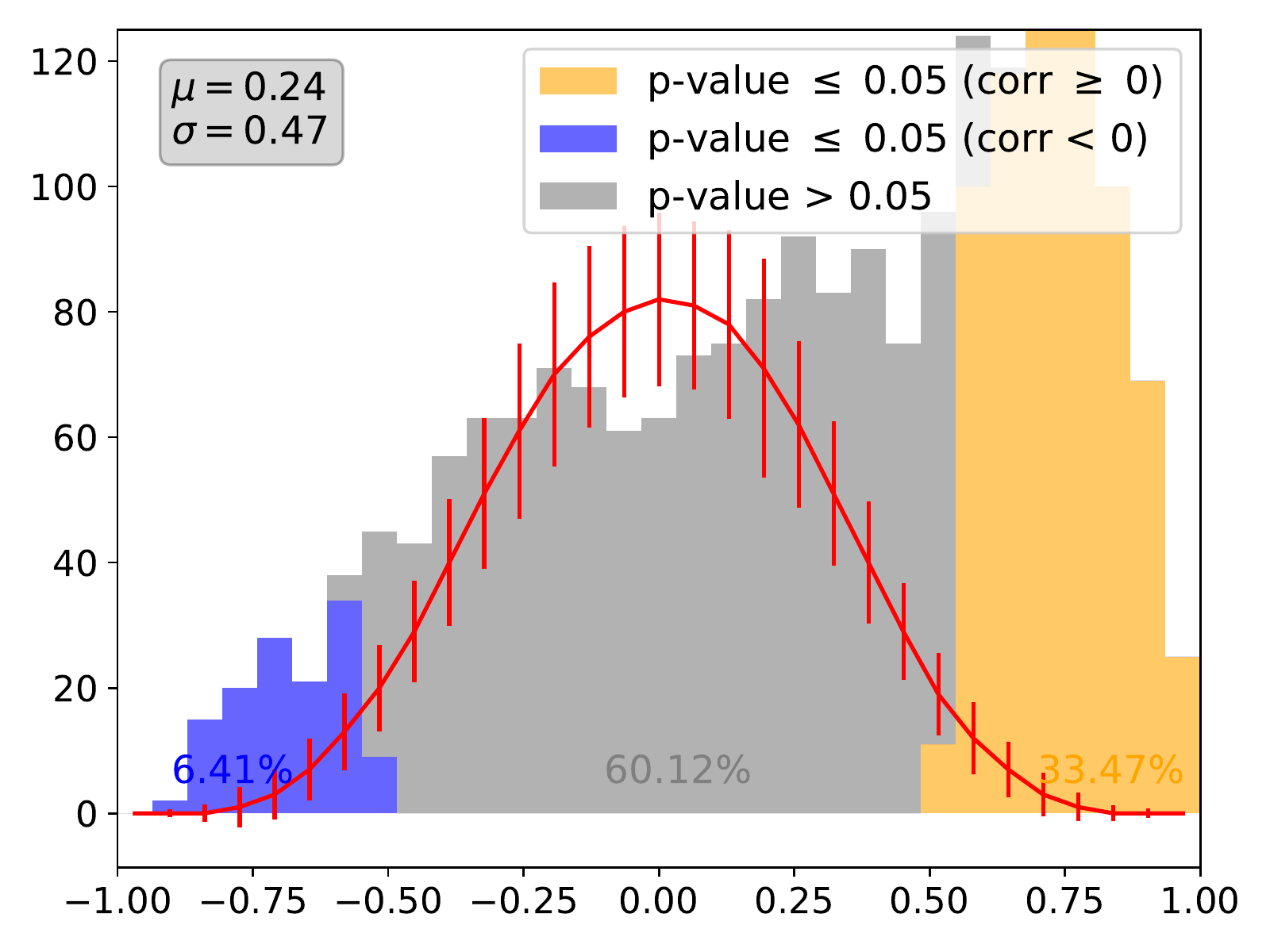}}
    \subfigure[]{\includegraphics[width=0.49\textwidth]{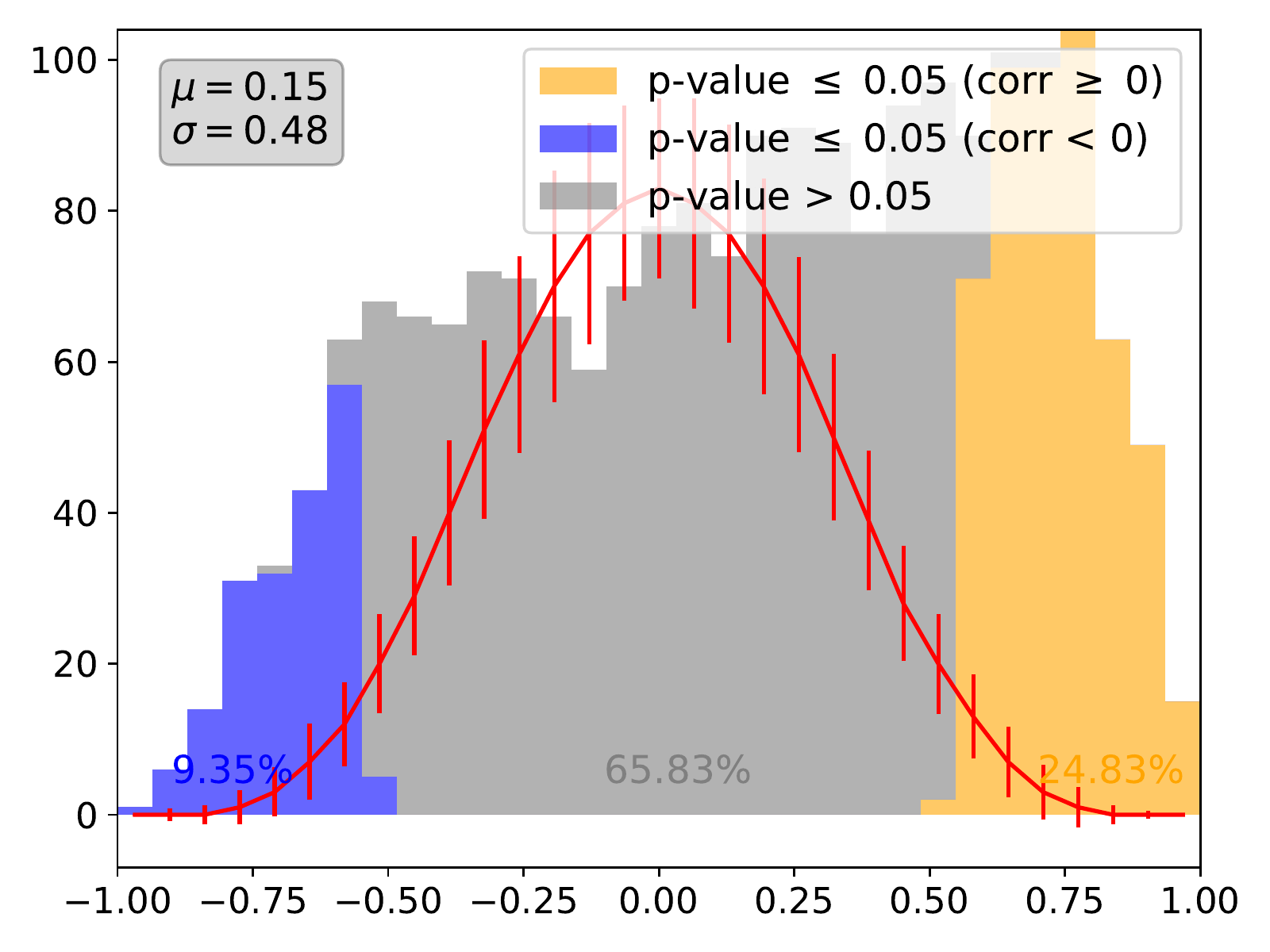}}
    \subfigure[]{\includegraphics[width=0.49\textwidth]{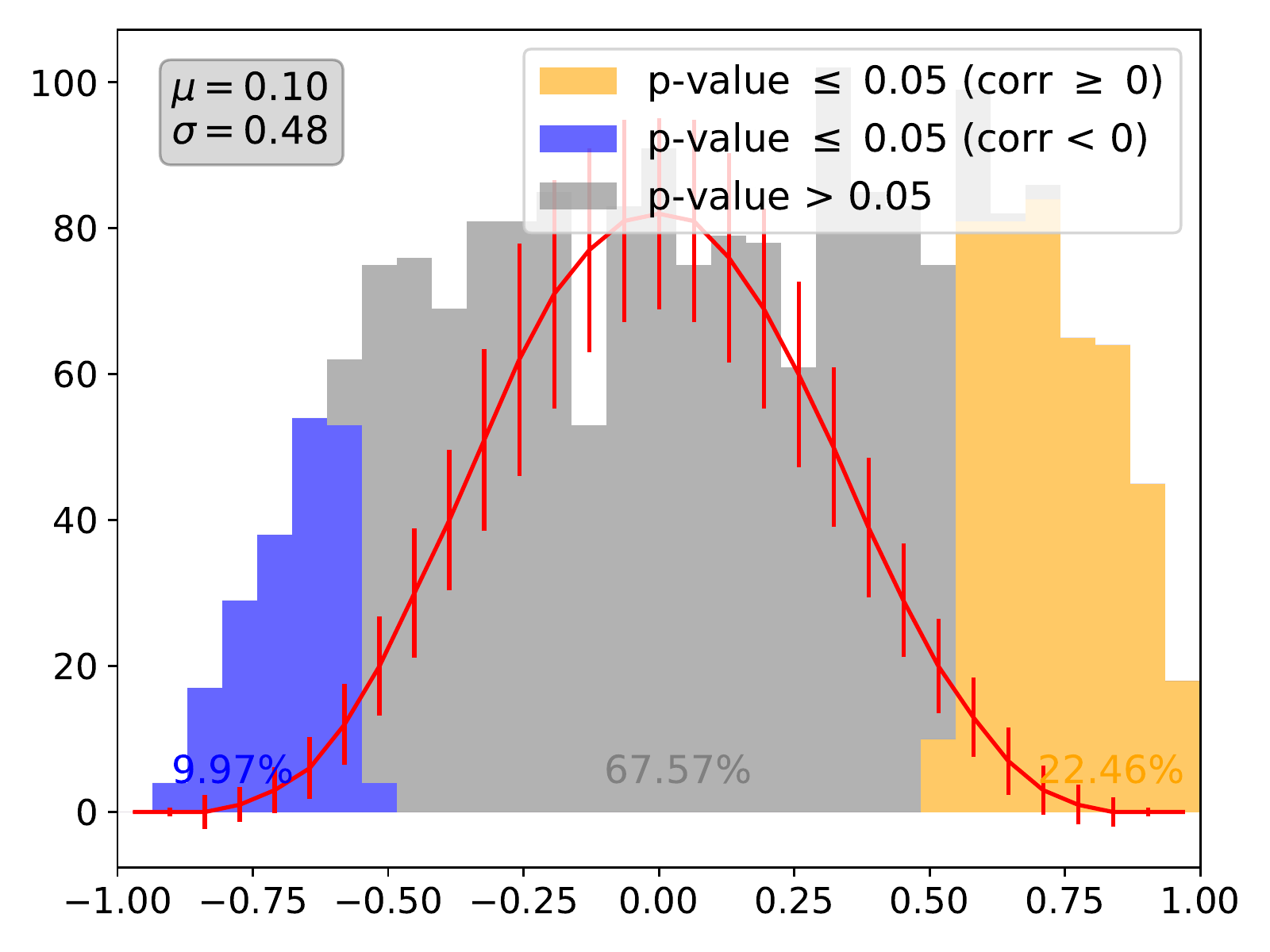}}
    \subfigure[]{\includegraphics[width=0.49\textwidth]{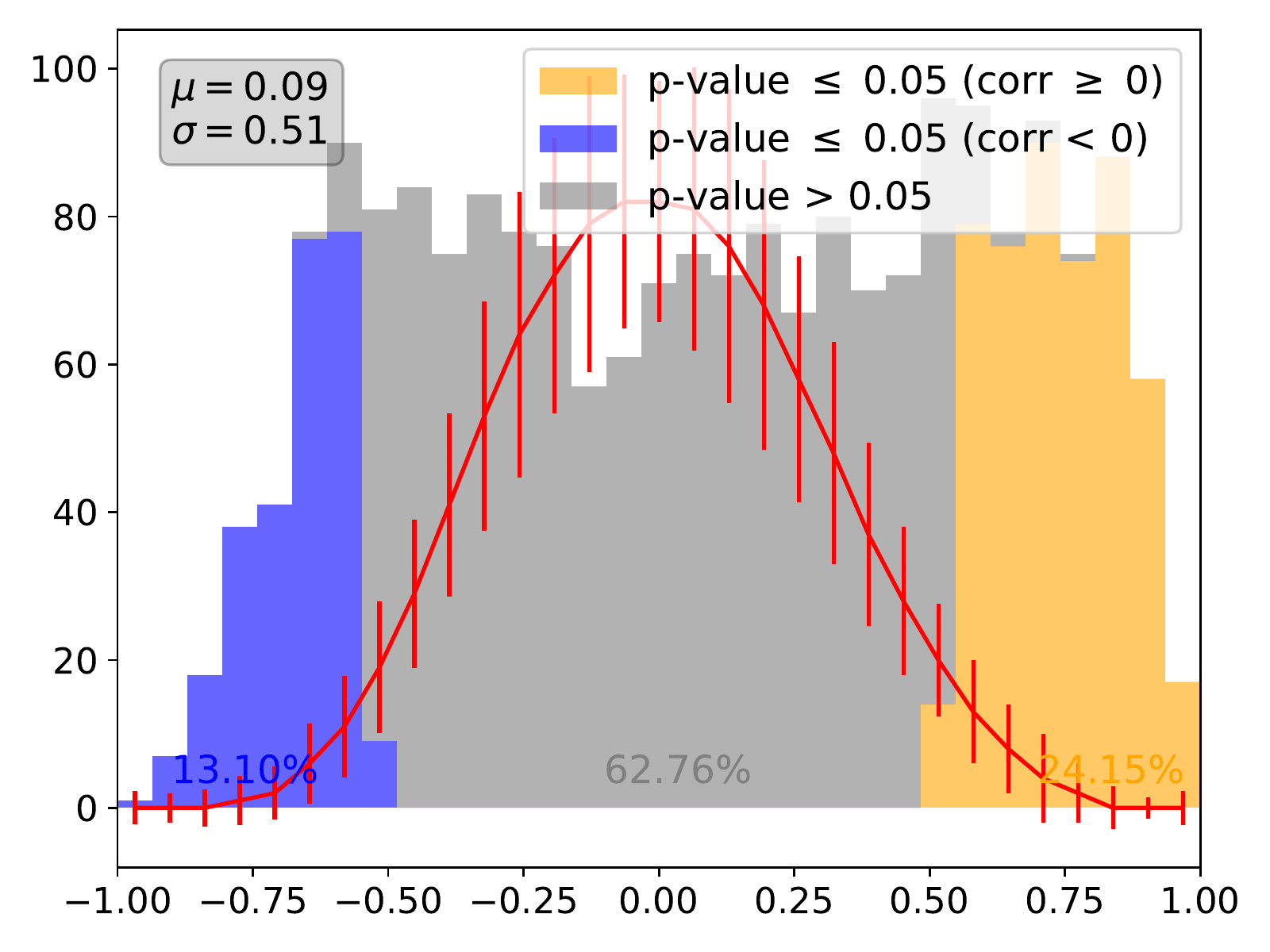}}
    \caption{Correlation between \emph{diversity of references} and \emph{citations per paper}. Panels (a)-(d) correspond to quartiles of authors sorted, in increasing order, by number of publications. The distribution of correlations obtained with the adopted null model is shown in the red curve. }
    \label{fig:divout-cit}
\end{figure}

In Figure \ref{fig:divout-cit}, we show the histograms of correlations between the \emph{diversity of references} and the \emph{number of citations per paper}. A stronger positive correlation is observed specially for authors with lower productivity. In panel (a), one-third of authors displayed a positive correlation between references diversity and visibility, while in (b), the same behavior occurred for one-fourth of all authors.
In both cases, positive correlations are more frequent than negative correlations. We found, $q = 5.22$ and $q = 2.65$, respectively for authors in classes (a) and (b). Authors in classes (c) and (d) displayed $q$ values similar to those observed in class (b).

The analysis of reference diversity showed that the way in which authors cite other works may affect their visibility in the near future. This effect was found to be more relevant than the productivity since significant positive correlations were found in up to 25\% of authors. This effect might be related to the fact that diverse references might attract attention from other subfields, favoring thus the dissemination of authors' visibility in other scientific communities. In fact, a similar effect has been reported at the journal analyses comparing the relationship between journals impact factor and interdisciplinary indexes~\cite{silva2013quantifying}. Similar effects have also been observed in diffusion systems, where the presence across different communities benefits the spreading of agents~\cite{kaiser2007criticality}. While it is not possible to establish a causal effect, our results suggest that references diversity (inside a field) might play a role in predicting authors' visibility.

\subsection{Correlations between reference features and citations diversity} \label{sec2}

While in the previous section we analyzed how references features correlate with visibility, here we investigate the relationship between references and the diversity of citations. Because citations diversity can be seen as an interdisciplinary index (see e.g.~\cite{silva2013quantifying}), this section analyzes how the choice (and quantity) of references is related to authors interdisciplinarity.

\begin{figure}[ht!]
    \subfigure[]{\includegraphics[width=0.48\textwidth]{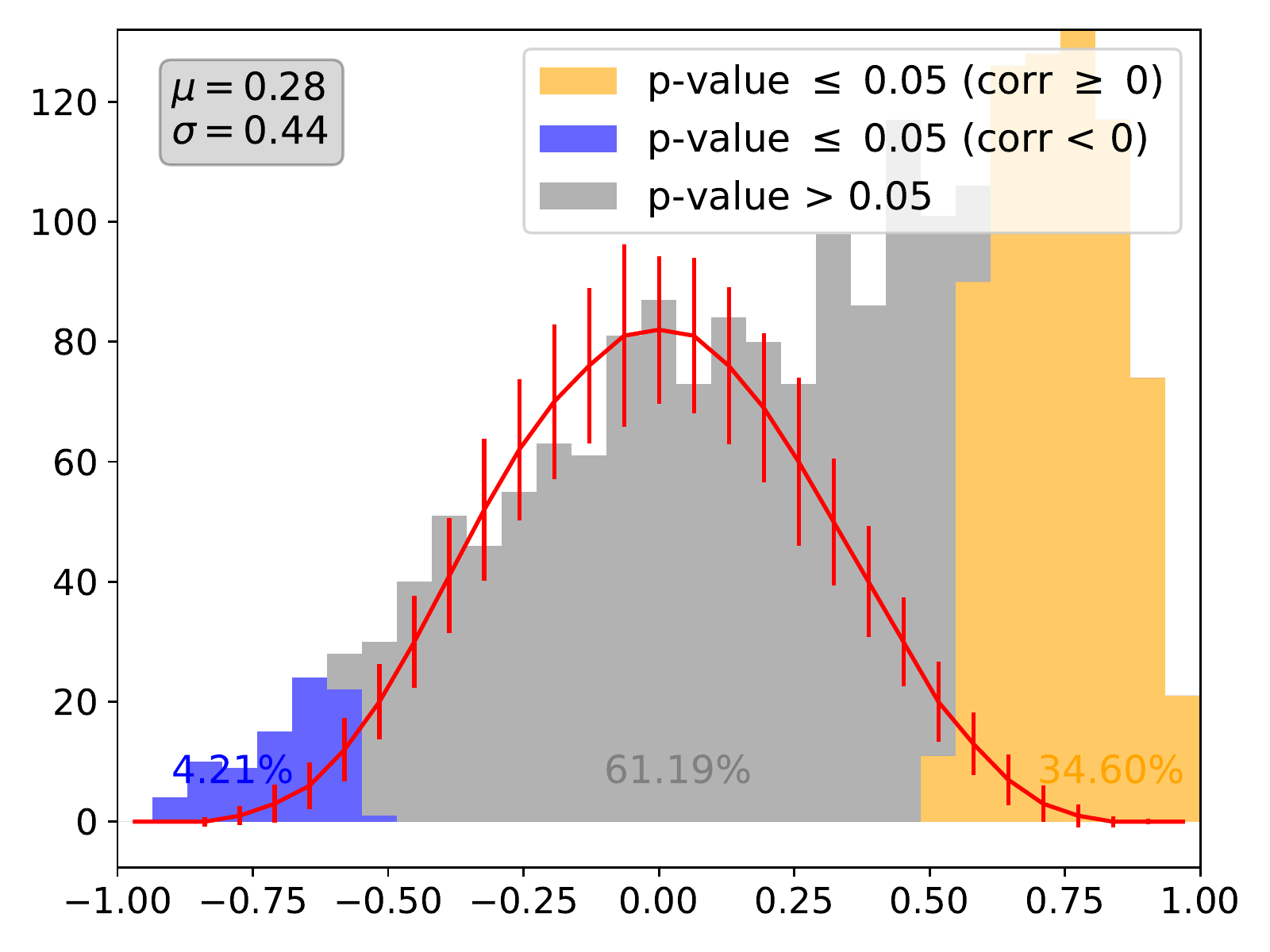}}
    \subfigure[]{\includegraphics[width=0.48\textwidth]{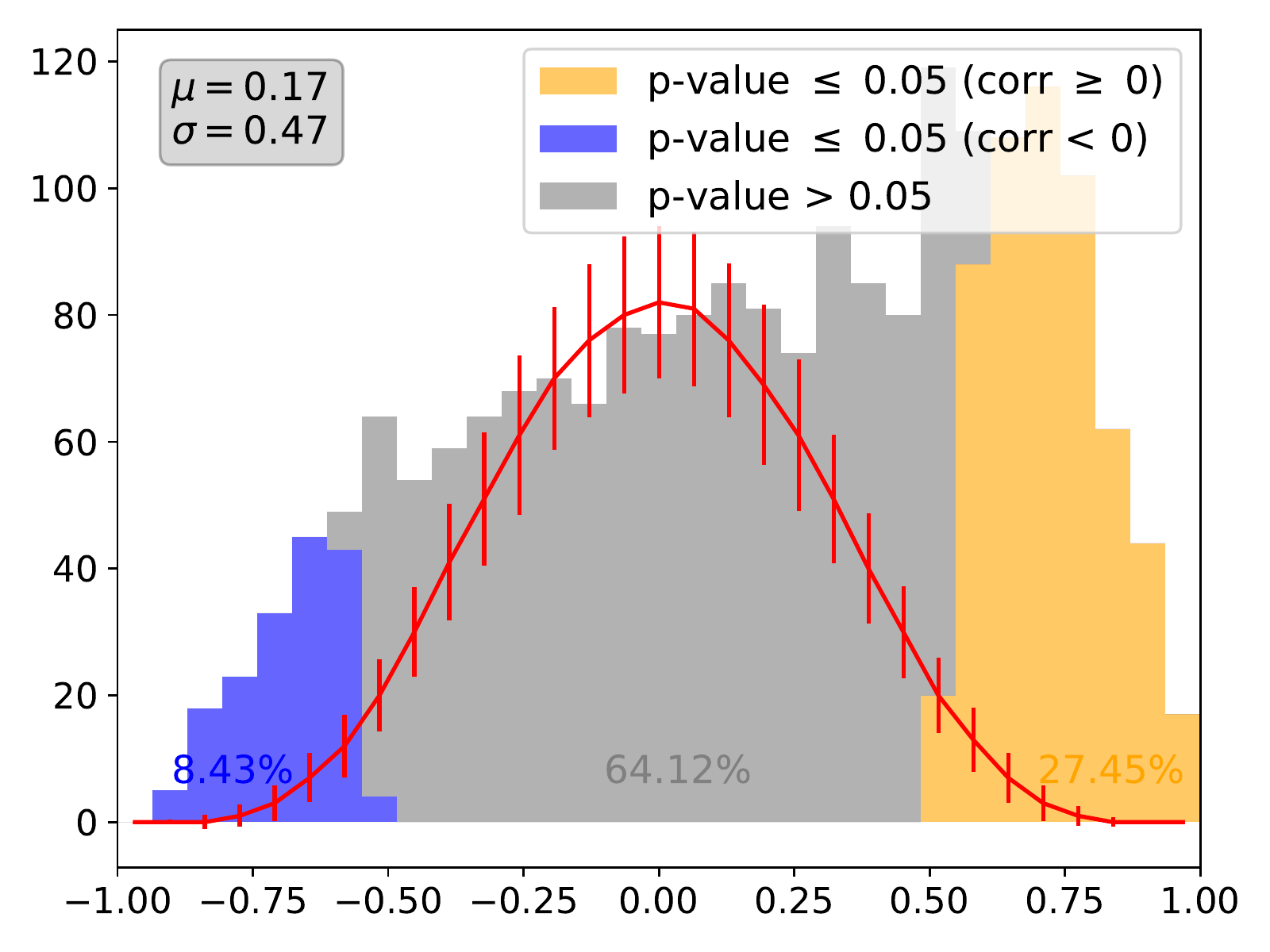}}
    \subfigure[]{\includegraphics[width=0.48\textwidth]{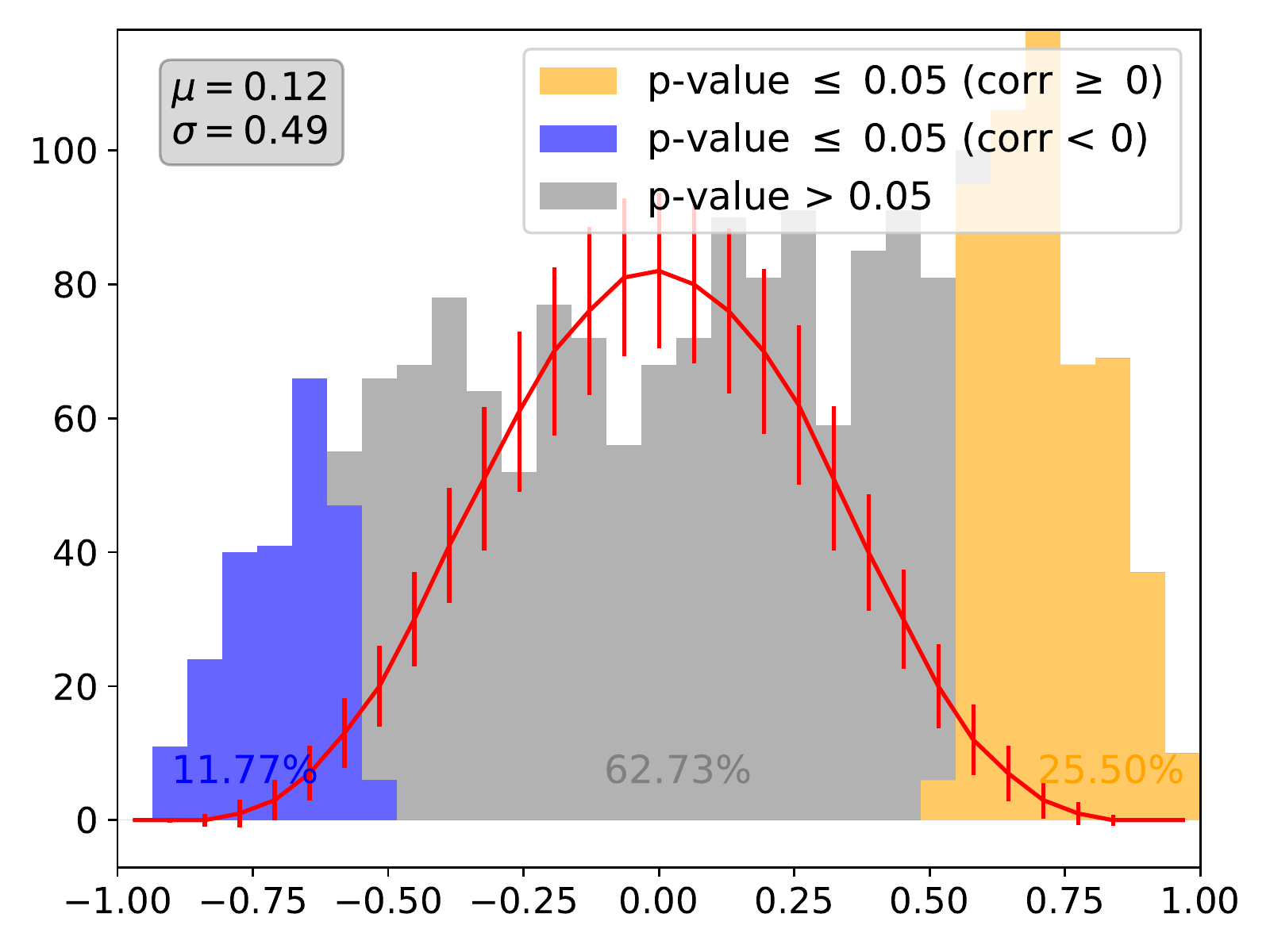}}
    \subfigure[]{\includegraphics[width=0.48\textwidth]{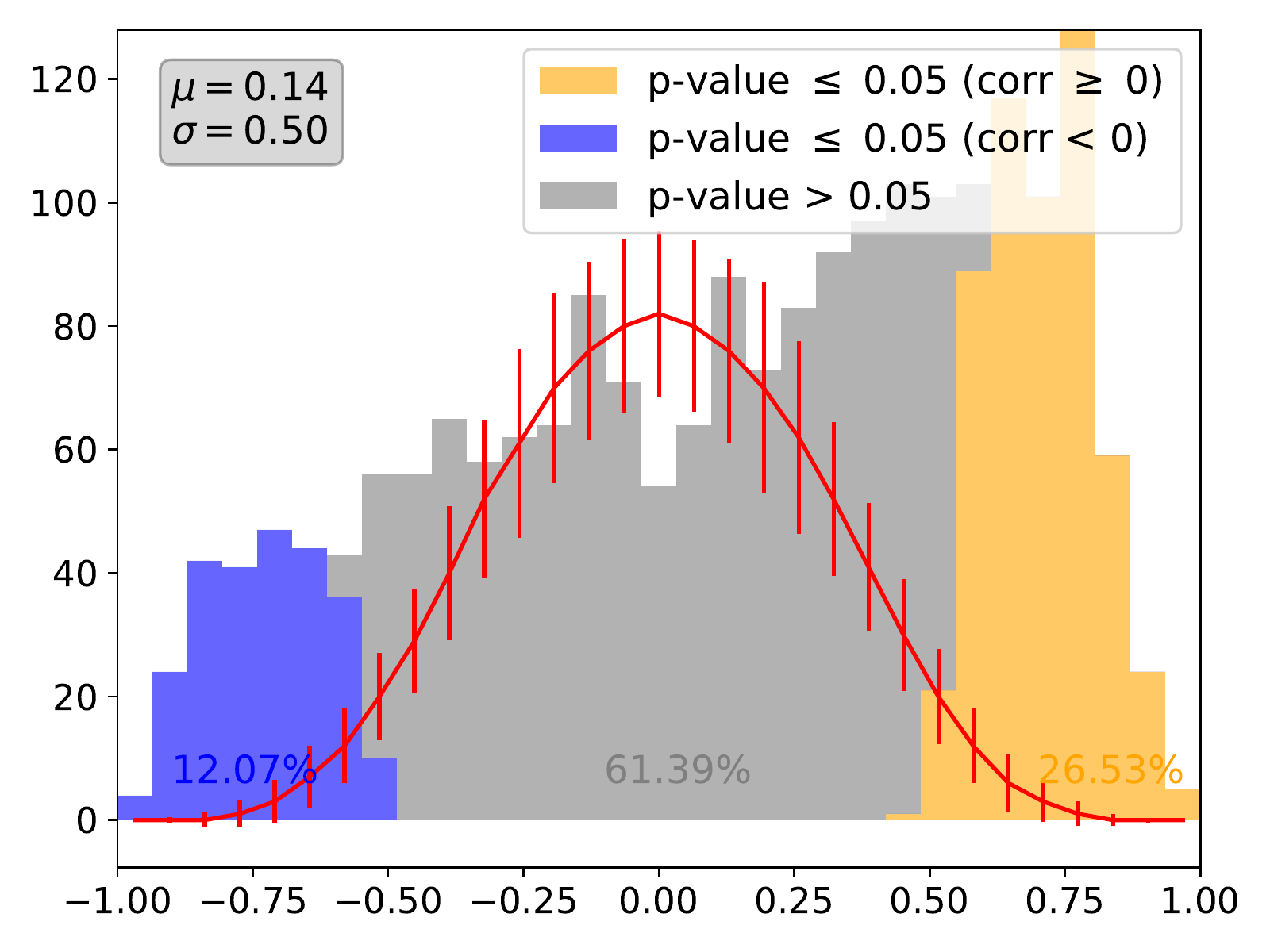}}
    \caption{Correlation between number of \emph{published papers} and \emph{citations diversity}. Panels (a)-(d) correspond to quartiles of authors sorted, in increasing order, by number of publications. The red curve denotes the distribution of correlations obtained with the adopted null model. }
    \label{fig:papers-divin}
\end{figure}

Figure \ref{fig:papers-divin} depicts the correlations between the \emph{number of published papers} and \emph{citation diversity}. As observed in the results reported in the previous section, for most of the authors there is no significant correlation between the considered variables. However, a positive correlation is observed for 1/3 of all authors in class (a), while 1/4 of authors displayed a positive correlation in the other classes. A negative correlation is less frequent than positive correlations. In addition, the values of $q$ decreases with productivity, since
we obtained $q_A = 8.2$, $q_B=3.3$, $q_C=2.2$ and $q_D=2.2$ respectively for classes (a), (b), (c) and (d). The results suggest, therefore, that an increase in productivity is more likely to play a role in increasing interdisciplinary for the class of authors with a lower degree of productivity.

\begin{figure}
    \centering
    \subfigure[]{\includegraphics[width=0.49\textwidth]{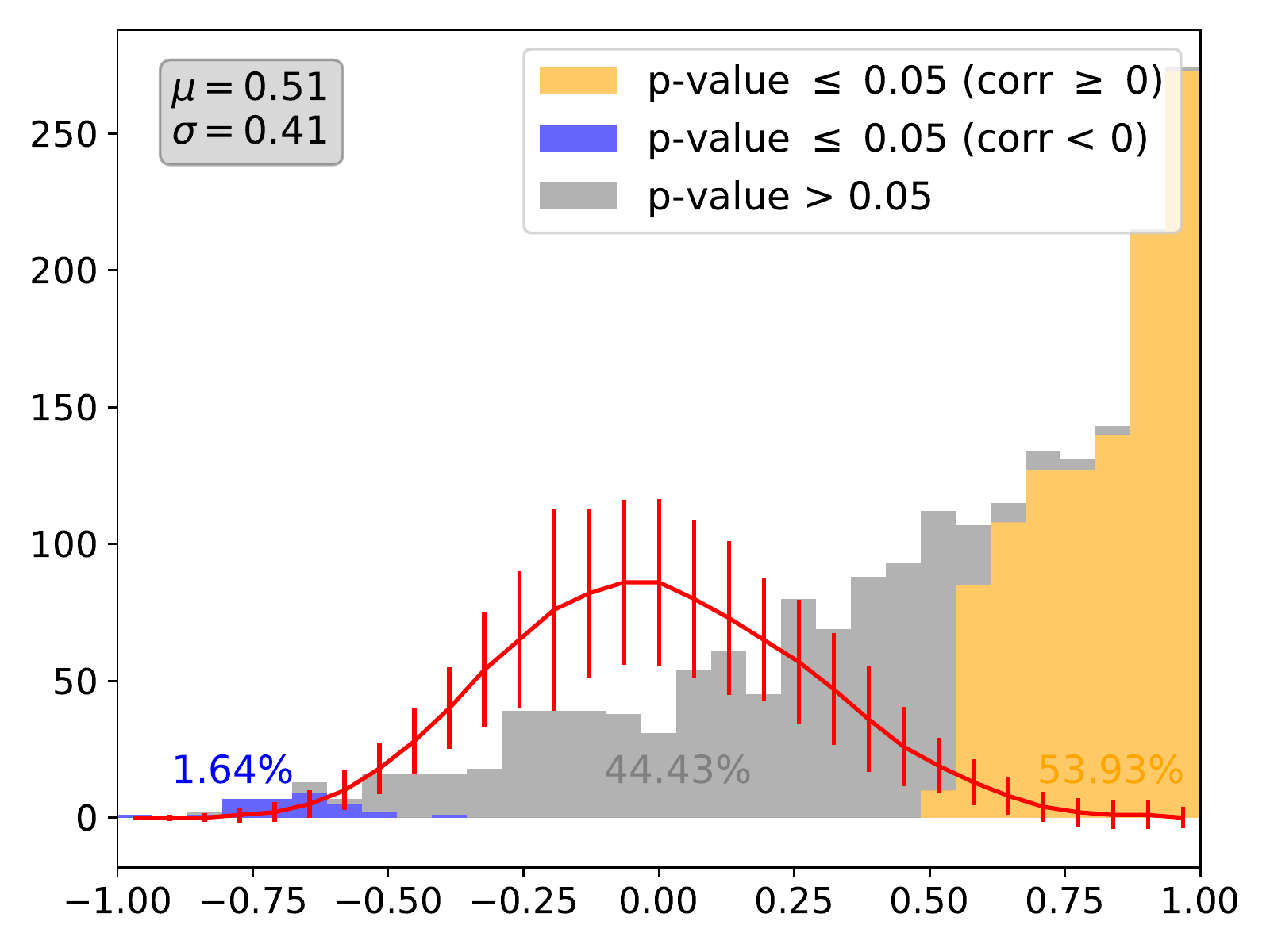}}
    \subfigure[]{\includegraphics[width=0.49\textwidth]{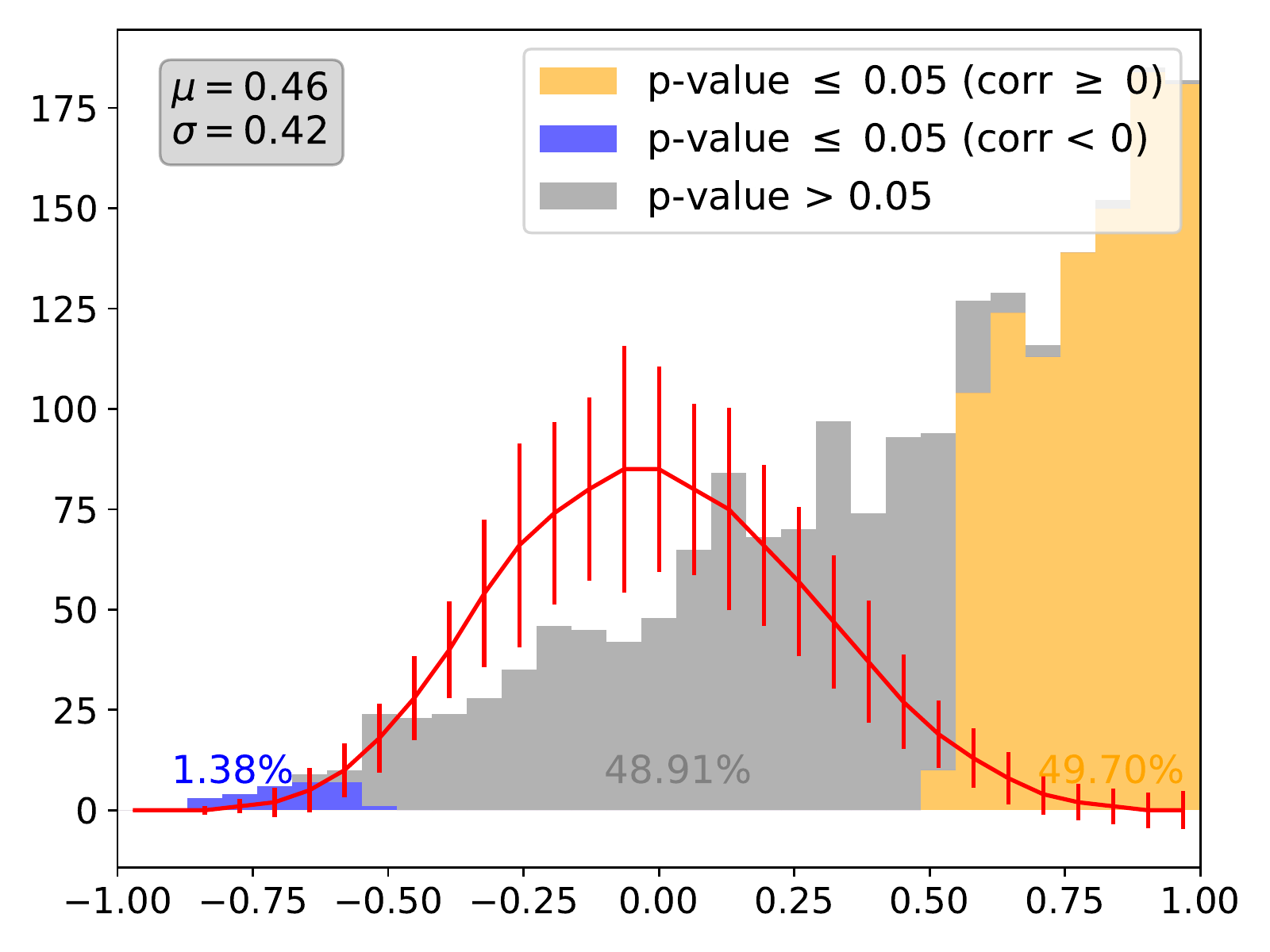}}
    \subfigure[]{\includegraphics[width=0.49\textwidth]{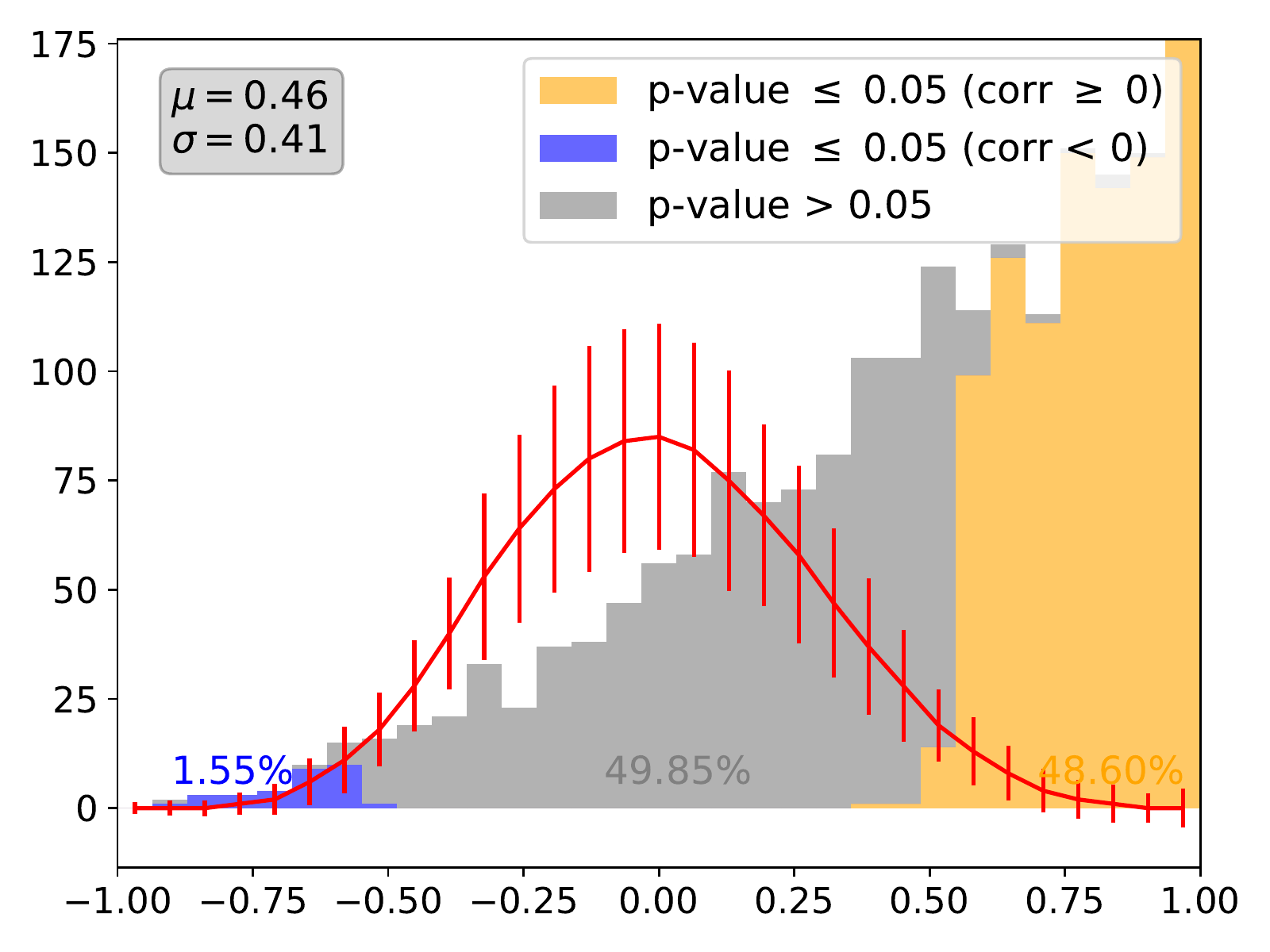}}
    \subfigure[]{\includegraphics[width=0.49\textwidth]{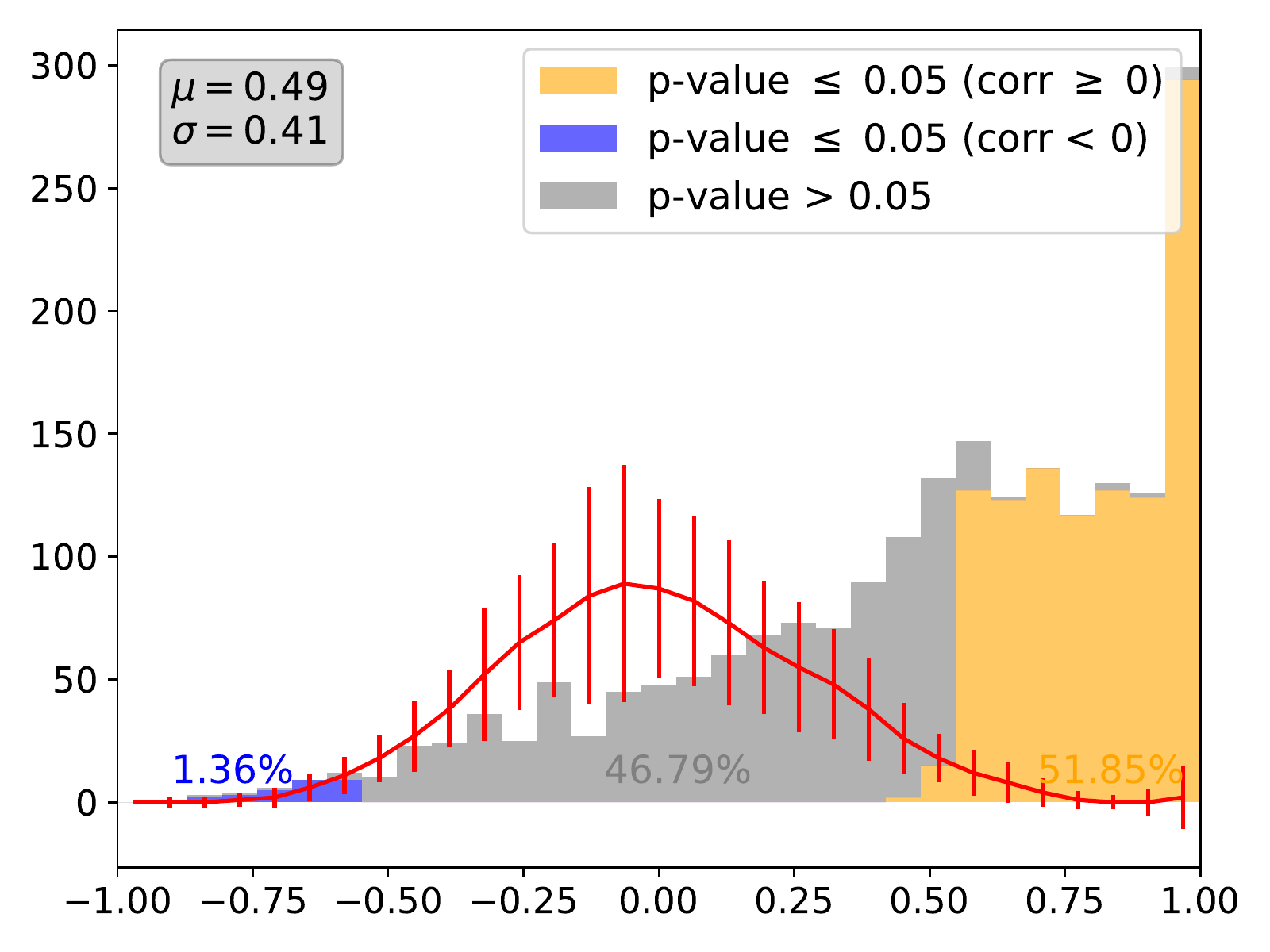}}
    \caption{
    Correlation analysis between \emph{reference diversity}  and \emph{citation diversity}. Panels (a)-(d) correspond to quartiles of authors sorted, in increasing order, by number of publications. The red curve denotes the distribution of correlations obtained with the adopted null model. }
    \label{fig:divout-divin}
\end{figure}

The association between references and citation diversity was also analyzed.
This result is shown in Figure \ref{fig:divout-divin}. The observed correlations are much stronger than the ones analyzed so far. The null model distribution is clearly not compatible with the real data. Here, a significant relationship between reference and citation diversity arises for \emph{more than 50\% of all authors}. Surprisingly, virtually all significant correlations are positive. The percentage of positive correlations reaches roughly 50\%, while significant negative correlations were observed for roughly 1.5\% of all authors. Another distinctive feature of the relationship between reference and citation diversity lies in the fact that the relationship is similar for all classes of authors. This result is, therefore, a strong evidence that researchers who cite papers from many other disciplines might be also cited by many other subareas. In other words, if authors display a diverse behavior when citing other papers, they also tend to be cited by other diverse subareas. Because citation diversity can be seen as a way to measure authors' interdisciplinary~\cite{silva2013quantifying}, most of the authors adopting larger reference diversity in a given period are expected to increase their interdisciplinarity indexes in the near future.


\subsection{Interplay between interdisciplinarity and citations} \label{sec:interplay}

\begin{figure}[ht!]
    \centering
    \subfigure[]{\includegraphics[width=0.48\textwidth]{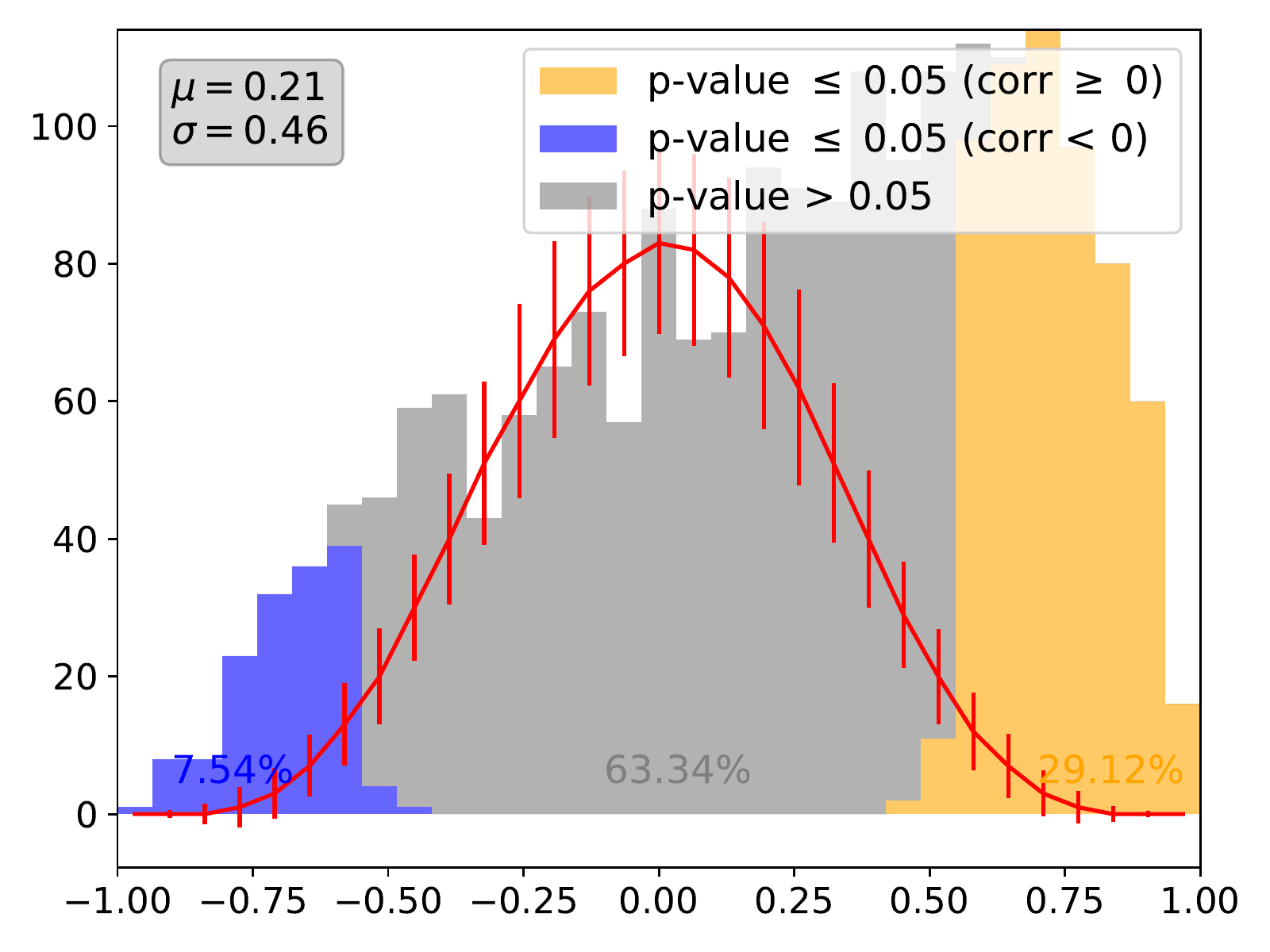}}
    \subfigure[]{\includegraphics[width=0.48\textwidth]{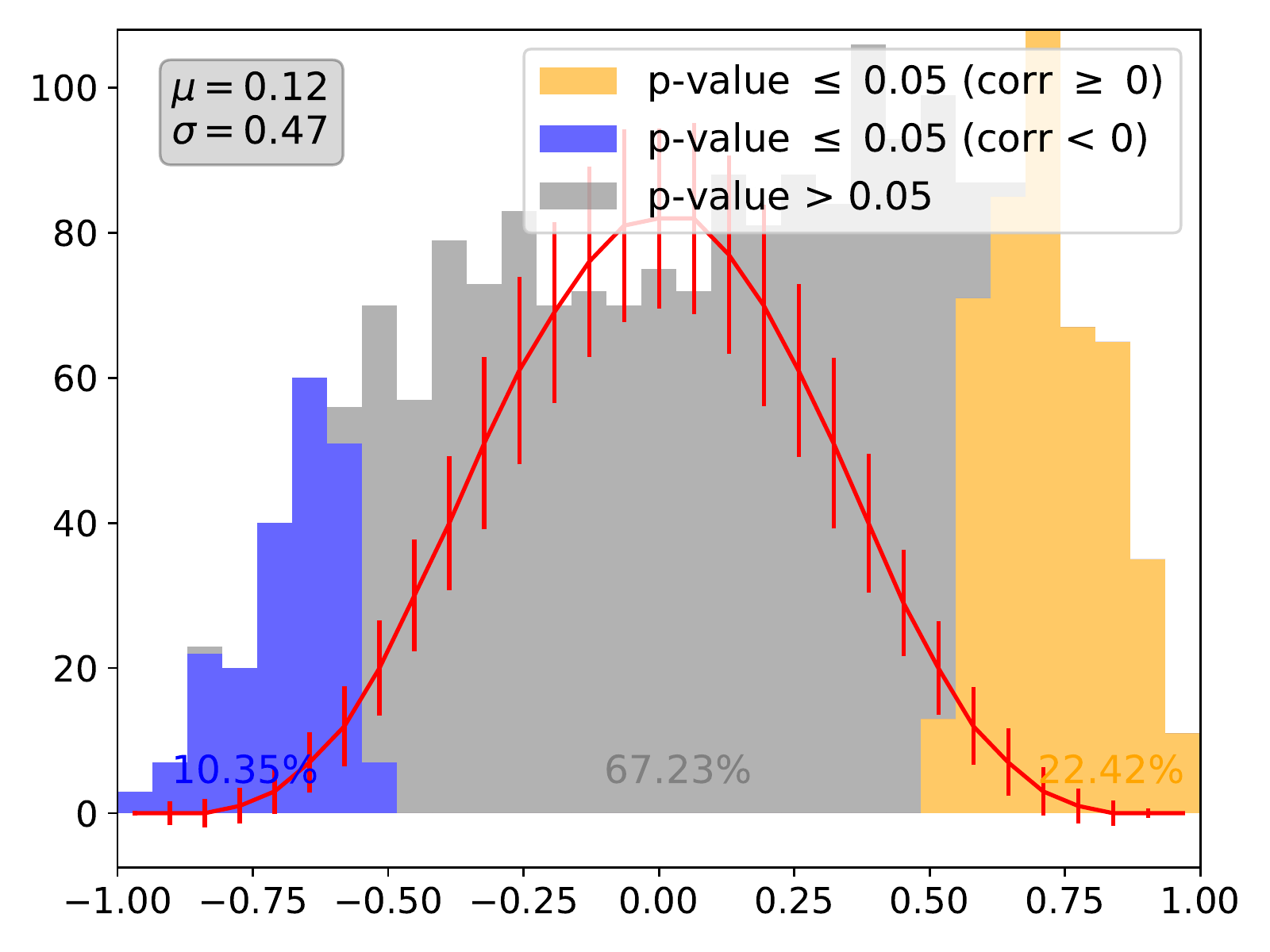}}
    \subfigure[]{\includegraphics[width=0.48\textwidth]{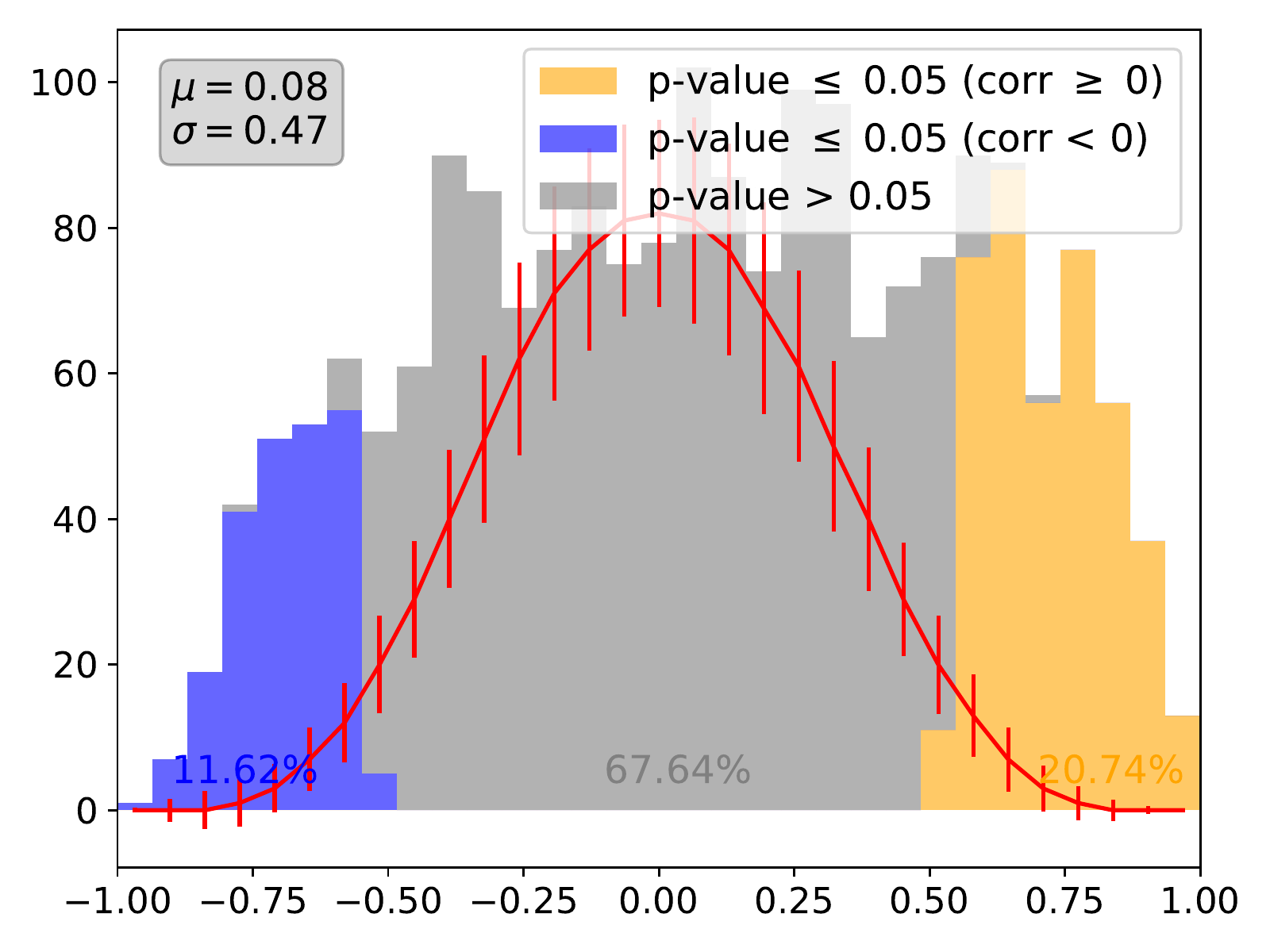}}
    \subfigure[]{\includegraphics[width=0.48\textwidth]{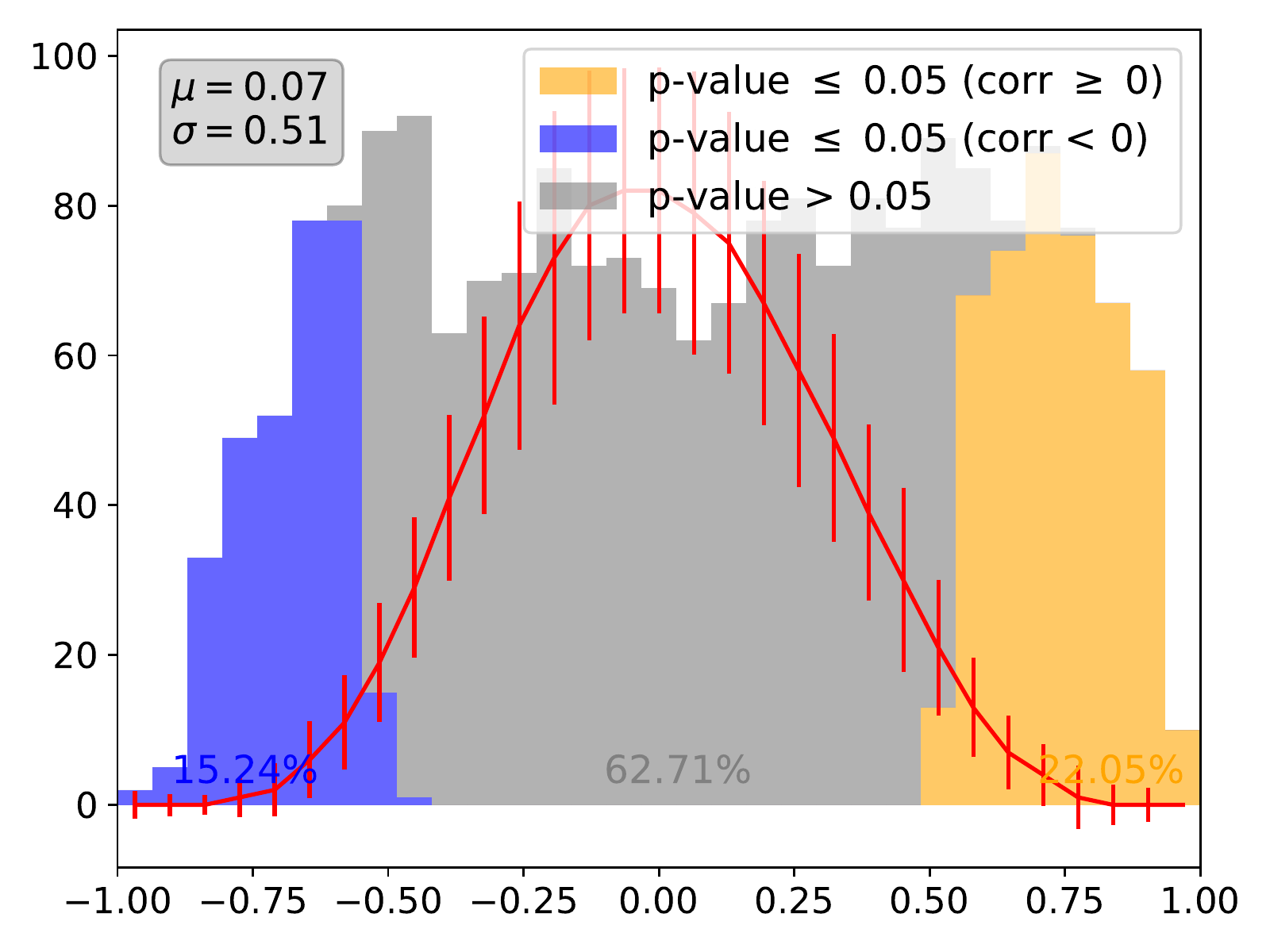}}
    \caption{Correlation analysis between \emph{citation diversity}  and \emph{citations per paper}. Panels (a)-(d) correspond to quartiles of authors sorted, in increasing order, by number of publications. The red curve denotes the distribution of correlations obtained with the adopted null model.
    }
    \label{fig:divin-cit}
\end{figure}

Here we analyze the relationship between interdisciplinarity of authors (computed as citation diversity) and the number of citations. While  studies have shown that a positive correlation exists between journals interdisciplinarity and impact factor~\cite{silva2013quantifying}, only a few studies have touched on this issue at the author level. Here we found that for most of the authors ($\geq 62\%$), there is no significant relationship between interdisciplinarity and the number of citation received by the authors. However, a positive correlation can be found for a considerable fraction of authors. This is more evident again for authors in class (a), as shown in Figure \ref{fig:divin-cit}: roughly $30\%$ displayed a significant positive correlation. Figure \ref{fig:divin-cit} also reveals a tendency of higher positive significant correlation over a negative one: the values of $q$ for each class are  $q_A=3.9$, $q_B=2.2$, $q_C=1.8$ and $q_D=1.4$. As observed in other associations studied here, higher values of $q$ are found for authors in the group of lower productivity.

\begin{figure}[ht!]
    \centering
    \subfigure[]{\includegraphics[width=0.48\textwidth]{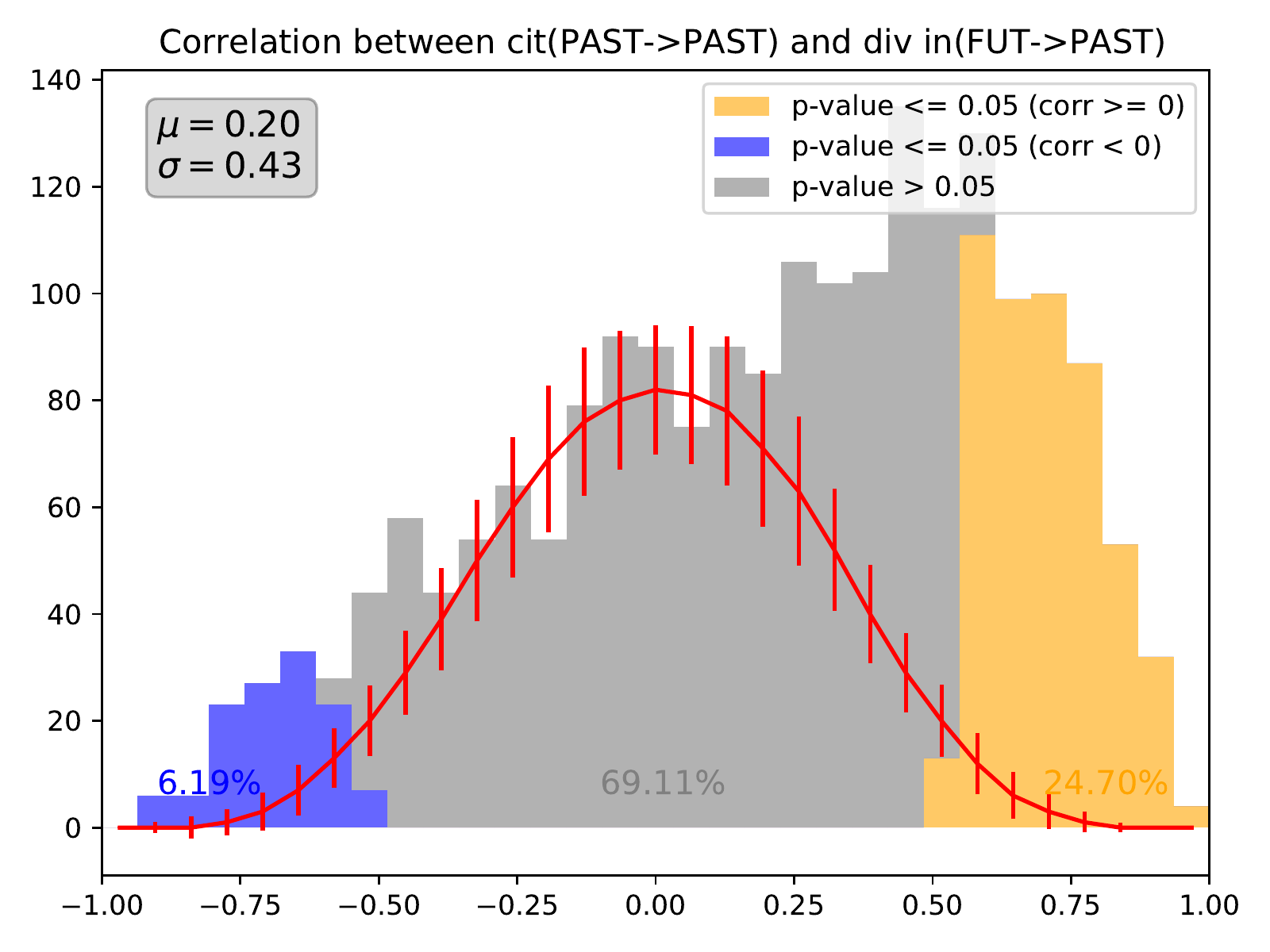}}
    \subfigure[]{\includegraphics[width=0.48\textwidth]{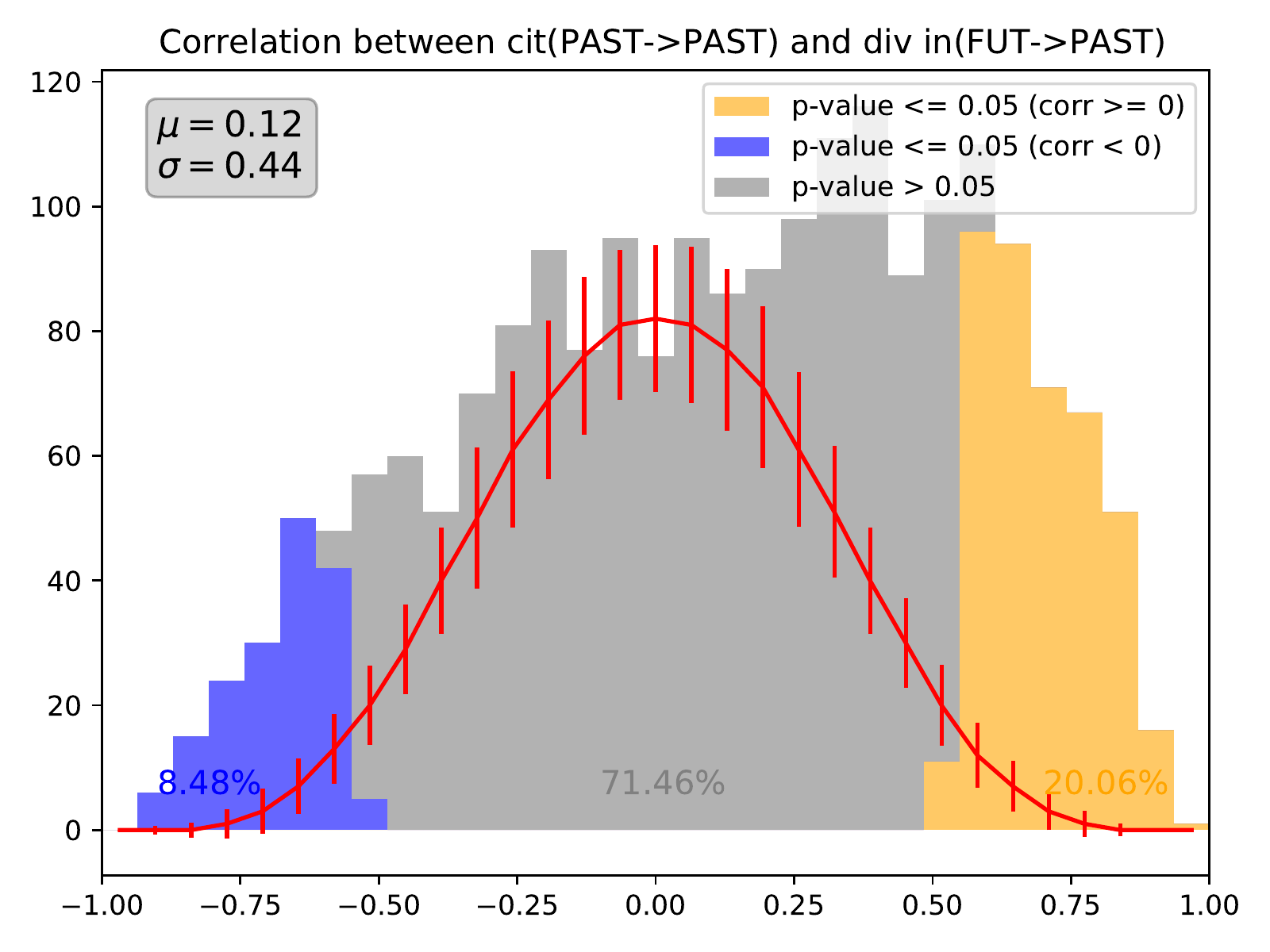}}
    \subfigure[]{\includegraphics[width=0.48\textwidth]{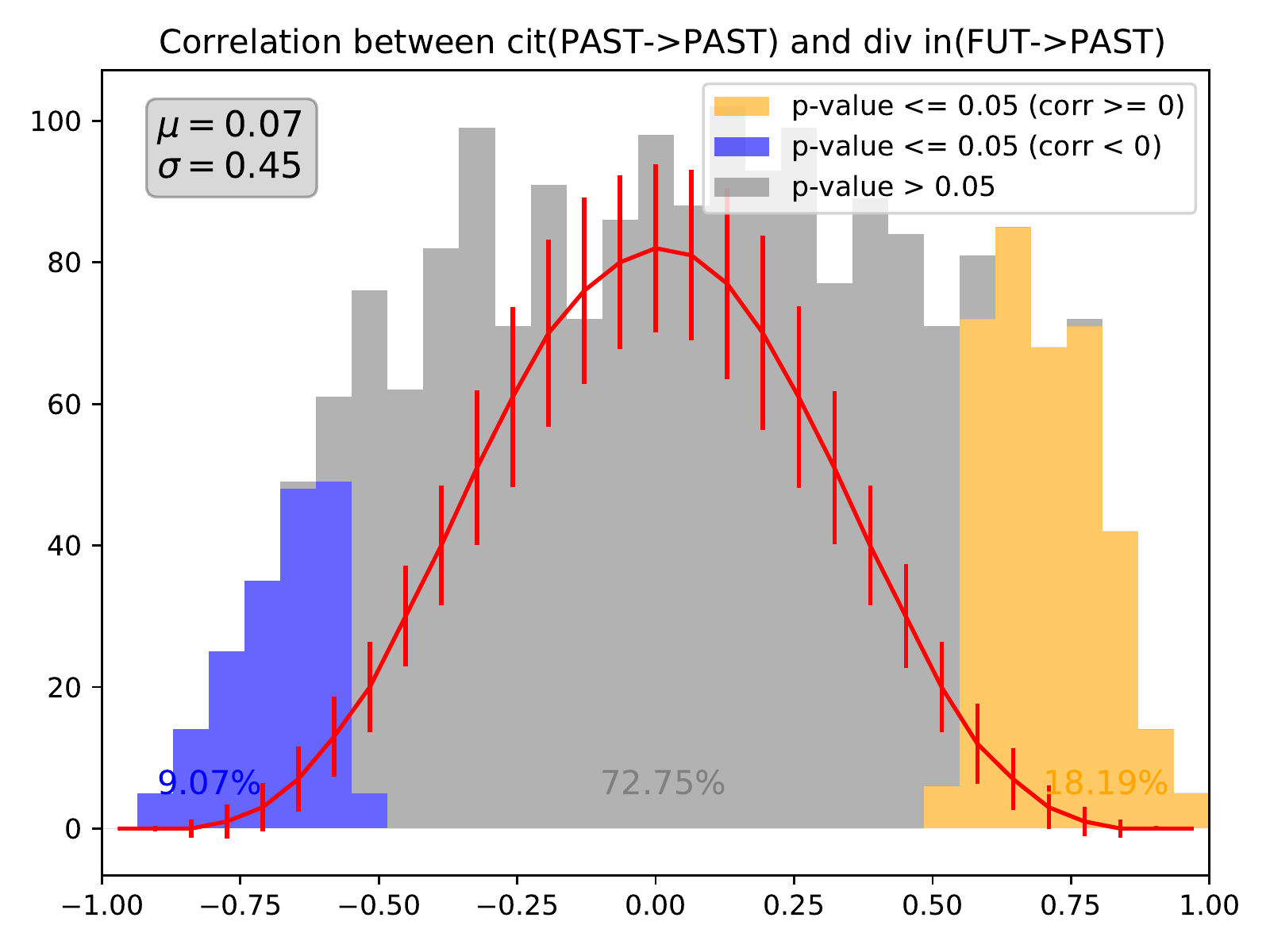}}
    \subfigure[]{\includegraphics[width=0.48\textwidth]{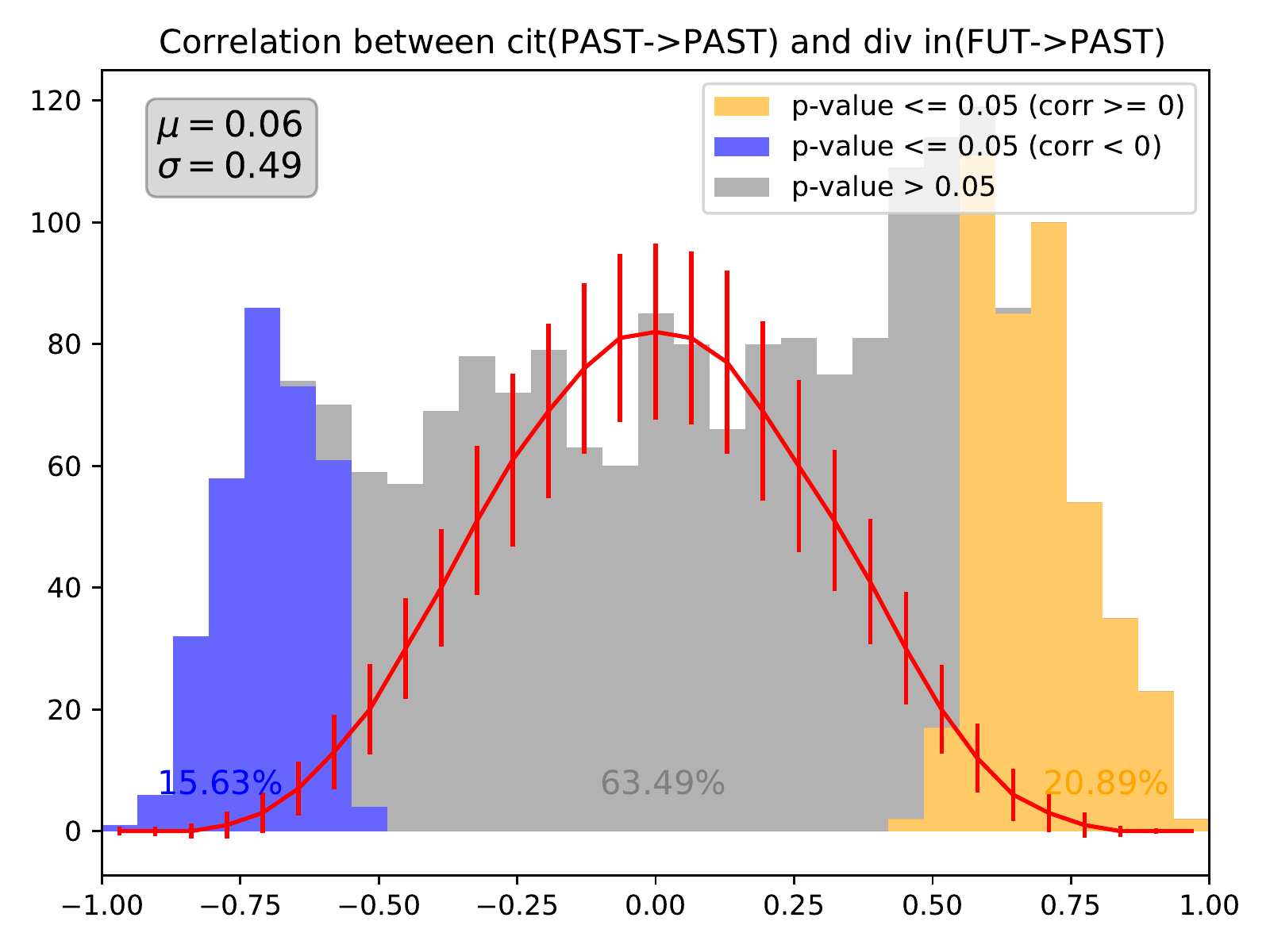}}
    \caption{Correlation analysis between \emph{citations per paper} and \emph{citation diversity}. Panels (a)-(d) correspond to quartiles of authors sorted, in increasing order, by number of publications. The red curve denotes the distribution of correlations obtained with the adopted null model. }
    \label{fig:cit-divin}
\end{figure}

While Figure \ref{fig:divin-cit} only show the relationship between interdisciplinarity and future visibility, it would be still interesting to see if there is an inverse effect.
To investigate if variation in citations is correlated to a future variation in interdisciplinarity we conducted an analysis similar to the one provided in Figure \ref{fig:divin-cit}. The histograms of correlations are shown in Figure \ref{fig:cit-divin}. Overall the histograms are similar to the ones depicted in Figure \ref{fig:divin-cit}, but here the fraction of significant positive correlations are smaller. This is evident e.g. for authors in (a): the fraction of positive correlations drop from $29.1\%$ to $24.7\%$. This suggests that a variability in citation counts is weakly correlated with the future author interdisciplinarity for most of authors. In other words, for most of the authors, a variability in visibility does not affect the future authors' interdisciplinary indexes.

\section{Conclusion}

In this paper, we analyzed whether relevant scholarly variables are correlated. We proposed a framework to probe if features extracted from authors' recent history are correlated with metrics observed a few years later.
While some correlations are trivial and were not object of study (such as correlations between citations in subsequent time periods~\cite{recency}), we studied the correlation between other variables of interest. We focused our analysis on simple, yet relevant metrics, including number of publications, number of citations, references diversity and authors' interdisciplinarity (measured via citation diversity).

Several interesting results have been obtained. Among the associations studied, we found that the strongest correlations were obtained between references diversity and authors' interdisciplinarity. Here we found a reciprocal tendency: if authors increase their diversity when citing other papers, received citations will also tend to increase. This pattern was observed for more than 50\% of authors. The relationship between productivity and visibility was found to be more prominent for authors with a lower productivity. While no significant correlation exists for most of authors, about 20\% showed a positive and significant correlation. A stronger association was obtained when analyzing the relationship between references diversity and future citation. For the class of authors with a lower productivity, we found that roughly 1/3 of authors displayed a significant positive correlation between references diversity and visibility. We also studied the association between references and citation diversity and found out that the fraction of positive significant correlations ranges between 18-30\% across different classes of authors.

Our study shed lights into the relationship between current and future researchers' activity. The results obtained here could be extended in diverse studies to provide mechanisms to predict authors' behavior, given the recent researchers' history. Future research could dive into other research questions arising from our analysis. For example, while we found that significant positive correlations are more likely to happen than negative ones, it would be interesting to probe which factors make authors display opposite behaviors for the same variables of interest. Another interesting feature that could be studied concerns the causality of the obtained correlations. Finally, a systematic study could be performed in different areas to check whether correlations are more significant in specific subfields.




\section*{Acknowledgments}

D.R.A. acknowledges financial support from S\~ao Paulo Research Foundation (FAPESP Grant no. 2020/06271-0) and CNPq-Brazil (Grant no. 304026/2018-2). This study was financed in part by the Coordenação de Aperfeiçoamento de Pessoal de Nível Superior -- Brasil (CAPES) -- Finance Code 001.

\bibliographystyle{ieeetr}
\bibliographystyle{abbrv}







\end{document}